\documentclass[10pt,journal]{IEEEtran}
\usepackage{amsmath,amsfonts}
\usepackage{algorithm}
\usepackage{algorithmicx}
\usepackage{algpseudocode}
\usepackage{array}
\usepackage[caption=false,font=footnotesize,labelfont=sf,textfont=sf]{subfig}
\usepackage{textcomp}
\usepackage{stfloats}
\usepackage{url}
\usepackage{graphicx}
\usepackage{cite}
\usepackage{color} 
\usepackage{hyperref}
\hypersetup{pdfborder={0 0 0}}

\begin{document}

\title{Task-Oriented Feature Compression for Multimodal Understanding via Device-Edge Co-Inference}

\author{{Cheng Yuan,
	Zhening Liu,
        Jiashu Lv,
        Jiawei Shao,~\IEEEmembership{Member,~IEEE},
        Yufei Jiang,~\IEEEmembership{Member,~IEEE}, \\
	Jun Zhang,~\IEEEmembership{Fellow,~IEEE},
        and Xuelong Li,~\IEEEmembership{Fellow,~IEEE}
}

\thanks{C. Yuan, J. Shao and X. Li are with the Institute of Artificial Intelligence (TeleAI) of China Telecom, China (E-mail: 23s052018@stu.hit.edu.cn, shaojw2@chinatelecom.cn, xuelong\_li@ieee.org). C. Yuan and Y. Jiang are with the School of Electronic and Information Engineering, Harbin Institute of Technology, Shenzhen, China (E-mail: jiangyufei@hit.edu.cn). Z. Liu and J. Zhang are with the Department of Electronic and Computer Engineering, Hong Kong University of Science and Technology, Hong Kong (E-mail: zhening.liu@connect.ust.hk, eejzhang@ust.hk). J. Lv is with the School of Software and Microelectronics, Peking University, China (E-mail: 2021303044@mail.nwpu.edu.cn). 
The corresponding author is X. Li.
}}



\maketitle

\begin{abstract}
With the rapid development of large multimodal models (LMMs), multimodal understanding applications are emerging.
As most LMM inference requests originate from edge devices with limited computational capabilities, the predominant inference pipeline involves directly forwarding the input data to an edge server which handles all computations.
However, this approach introduces high transmission latency due to limited uplink bandwidth of edge devices and significant computation latency caused by the prohibitive number of visual tokens, thus hindering delay-sensitive tasks and degrading user experience.
To address this challenge, we propose a task-oriented feature compression (TOFC) method for multimodal understanding in a device-edge co-inference framework, where visual features are merged by clustering and encoded by a learnable and selective entropy model before feature projection.
Specifically, we employ density peaks clustering based on $K$ nearest neighbors to reduce the number of visual features, thereby minimizing both data transmission and computational complexity.
Subsequently, a learnable entropy model with hyperprior is utilized to encode and decode merged features, further reducing transmission overhead.
To enhance compression efficiency, multiple entropy models are adaptively selected based on the characteristics of the visual features, enabling a more accurate estimation of the probability distribution.
Comprehensive experiments on seven visual question answering benchmarks validate the effectiveness of the proposed TOFC method.
Results show that TOFC achieves up to 52\% reduction in data transmission overhead and 63\% reduction in system latency while maintaining identical task performance, compared with neural compression ELIC.
\end{abstract}

\section{Introduction}
\begin{figure*}[!t]
    \centering
    \includegraphics[width=.9\linewidth]{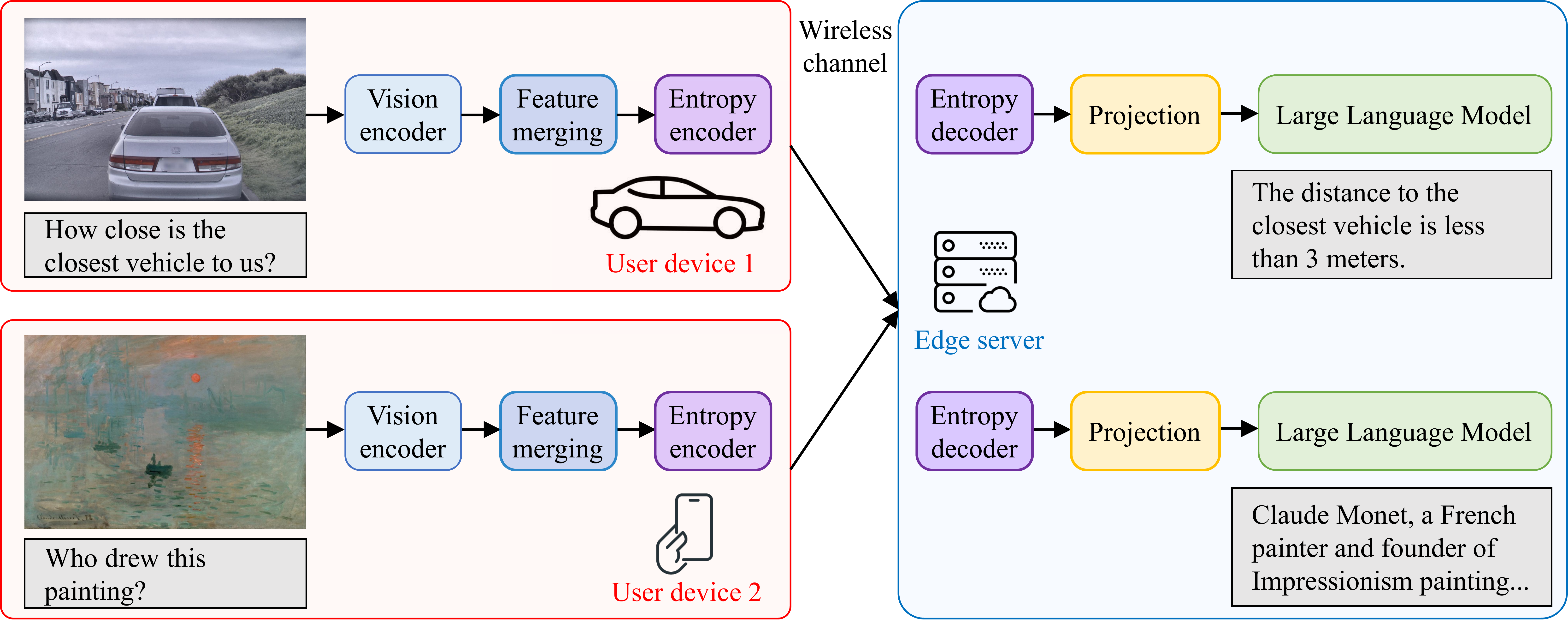}
    \caption{An example of the proposed device-edge co-inference system for multimodal understanding tasks, where the user device extracts and compresses visual features, and the intensive computation of LLM inference is handled by the edge server.
    }
    \label{fig_pipeline}
\end{figure*}
Large multimodal models (LMMs) have recently emerged as a powerful tool by integrating multimodal understanding capabilities with versatile large language models (LLMs) \cite{LMMsurvey1, LMMsurvey2}. 
LMMs have the potential to be applied to a variety of domains, including embodied artificial intelligence \cite{LMMembodied}, smart manufacturing, autonomous driving \cite{LMMdriving}, and personal assistants \cite{LMMagent}.
The architecture of mainstream LMMs for vision input comprises three components: a vision encoder, a feature projector, and an LLM \cite{llava, llava-ov, qwen2.5vl}.
The vision encoder is typically based on contrastive language–image pretraining (CLIP) \cite{CLIP}, which transforms image patches into visual features with lingual knowledge, benefiting from large-scale pretraining on image-text pairs.
Subsequently, the feature projector maps the extracted features into visual tokens, thus bridging the gap between CLIP and the LLM.
Finally, the LLM processes both visual and textual tokens and generates a response in natural language.
This architecture leverages the prior knowledge of both CLIP and LLM, accumulated from extensive pretraining, thereby reducing the training cost \cite{llava}.
However, a fundamental limitation of CLIP-based vision encoders \cite{CLIP, SigLIP} is that they represent a small image patch by a large number of features, leading to a prohibitive number of visual tokens.
As the computational complexity of attention mechanisms, which are widely adopted in LLMs, grows quadratically with the length of input sequence, the auto-regressive inference of LLMs incurs significant latency.
This challenge is particularly critical for applications that require low-latency responses and real-time interactions, such as analyzing the surrounding environment for autonomous robots \cite{LMMembodied}.
Moreover, most inference requests originate from edge devices with limited computational capabilities.
Consequently, the predominant LMM inference pipeline involves directly transmitting the input data to a proximate edge server, which handles all computations.
However, in scenarios where only the wireless network connection is available for the user device, the communication channel is highly dynamic and the uplink bandwidth is often stringent.
The transmission of visual data over such channel incurs substantial latency, which hinders the execution of delay-sensitive tasks \cite{TOCsurvey1}.
Therefore, the data communication overhead and high inference latency constitute the main design challenge of LMM inference.

To reduce bandwidth consumption and response time in LMM inference, a device-edge co-inference framework \cite{aiflow,aiflow_perspectives} stands out as a promising solution.
One co-inference paradigm under heated discussion is task-oriented communication \cite{TOCsurvey1, TOCsurvey2, TOCsurvey3, TOCmagazineshao}, which designs the transmission strategies based on the characteristics of downstream tasks rather than recovering the original data at the edge server, thus representing a paradigm shift from traditional data-oriented approaches.
In task-oriented co-inference schemes, the user device extracts task-relevant information and transmits only the essential features to the edge server, which handles the extensive computation of model inference.
While the device-edge co-inference framework with task-oriented communication has been explored for several vision tasks \cite{shao2022task, xie2023robust, shao2024task}, these methods are not directly applicable to LMM inference.
The primary challenge lies in the fact that LMMs rely on the CLIP model, which is pretrained on large-scale datasets, to extract visual features necessary for diverse downstream tasks.
Given this dependency, it is impractical to train an alternative vision encoder with more compact visual features from scratch to accommodate the constrained uplink bandwidth.

Various visual data compression techniques can be applied to satisfy the constraints on data transmission, including traditional compression methods \cite{wallace1991jpeg, skodras2001jpeg, webp} and neural image compression methods \cite{Balle2017, Balle2018, Cheng2020, ELIC}.
These methods reduce transmission overhead by compressing the input image, albeit at the expense of degraded image quality upon reconstruction at the edge server.
However, as these compression methods are designed to recover the original image in the pixel space, they do not selectively discard redundant information irrelevant to downstream tasks.
Consequently, when the available bandwidth is highly constrained, these methods may distort key semantic information essential for LMM inference, thus impairing performance on downstream tasks.

Another line of research aims to accelerate LMM inference by reducing the number of visual tokens, thereby minimizing the end-to-end system latency.
These methods either prune unimportant visual tokens or merge multiple visual tokens into a single feature at different stages of LMM inference, including vision encoding \cite{LLaVA-PruMerge, VisionZip}, feature projection \cite{TokenPacker, SeTok, QueCC} and LLM generation \cite{FastV}.
Given that the dimension of visual features produced by the vision encoder significantly grows after being projected to the LLM latent space \cite{llava1.5, llava-ov}, the transmission of visual features should be conducted before MLP projection to minimize data size under identical token count.
As the majority of token reduction methods are performed after MLP projection, these methods \cite{TokenPacker, SeTok, QueCC, FastV} do not alleviate the data transmission overhead in a device-edge co-inference framework, which constitutes a key bottleneck in our study.
Other works \cite{LLaVA-PruMerge, VisionZip} still assume that the entire inference process is executed on a single device and do not encode the visual features efficiently to meet the data transmission constraints.

\subsection{Contributions}
As illustrated in Fig. \ref{fig_pipeline}, we propose a task-oriented feature compression (TOFC) method for multimodal understanding in a device-edge co-inference framework, where visual features are merged by clustering and encoded by a learnable and selective entropy model before feature projection.
The main contributions are summarized as follows:
\begin{itemize}
    \item To the best of our knowledge, this is the first study to investigate task-oriented feature compression for LMM inference in a device-edge co-inference framework.
    The proposed TOFC method minimizes both data transmission overhead and the computational complexity of LMM inference, contributing to a substantial reduction in end-to-end system latency, especially when the user device relies on wireless connections with limited uplink bandwidth.
    The reduction in system latency enables delay-sensitive tasks and enhances user experience.
    \item First, we introduce a feature merging module (FMM), which employs density peaks clustering based on $K$ nearest neighbors (DPC-KNN) to partition visual features into clusters.
    Subsequently, average pooling is applied to each cluster, reducing the number of visual features to only 1.1\% to 4.4\% of the original number.
    This reduction in visual tokens minimizes both data transmission overhead and the number of visual tokens processed by the LLM, thus alleviating GPU memory usage and computational load at the edge server.
    \item Second, we employ a learnable entropy model with hyperprior to encode and decode the merged features, further reducing data transmission overhead.
    The merged features are modeled by Laplacian distribution with the mean and scale derived from the hyperprior using a learned network.
    Both the merged features and the hyperprior are quantized and encoded by an entropy coder based on the estimated probability distribution, thereby obtaining a more compact bitstream representation while maintaining distortion acceptable for downstream tasks.
    \item Third, to improve the accuracy of the estimated probability distribution, multiple entropy models are adaptively selected based on the characteristics of the visual features.
    Each entropy model is specialized in encoding specific features, thus achieving a data rate closer to the actual entropy and improving the overall compression efficiency.
    \item The effectiveness of the proposed TOFC method is validated through comprehensive experiments on seven visual question answering (VQA) benchmarks. 
    Results show that our TOFC method achieves substantial reductions in both data transmission overhead and system latency.
    Notably, on the RealWorldQA benchmark, our TOFC method reduces data transmission overhead by up to 52\% and system latency by 61\% while maintaining identical task performance compared with neural compression ELIC \cite{ELIC}.
    In addition, case studies are conducted to intuitively demonstrate the advantages of our TOFC method, and an ablation study validates the effectiveness of the proposed FMM and entropy model.
\end{itemize}

\subsection{Organization}
The rest of this paper is organized as follows. 
Section II provides a detailed description of related works.
Section III introduces the system model, and Section IV elaborates on the proposed TOFC method.
Experiment results are presented and analyzed in Section V, and Section VI concludes the paper.

\section{Related Works}
\subsection{Task-Oriented Communication}
The increasing demand for multimodal services has stimulated intensive interactions between user devices and the edge server, resulting in a surge in multimodal data transmission and posing severe challenges to existing communication systems. The conventional data-oriented transmission paradigm focuses on the reliable transmission of every single bit of multimodal input, while overlooking the inherent redundancy in the data. 
To address this limitation, task-oriented communication \cite{TOCsurvey1,TOCsurvey2,TOCsurvey3,TOCmagazineshao} proposes to extract the task-related information, thereby reducing data transmission overhead while maintaining satisfactory performance on downstream tasks. This objective is achieved by adopting the information bottleneck (IB) principle \cite{tishby2000information,shao2020bottlenet++,shao2021learning}, which maximizes task performance while minimizing communication overhead. 
Several recent works have leveraged task-oriented communication to reduce transmission overhead by over an order of magnitude. For instance, the authors in \cite{TOCvideo} establish a temporal entropy model to minimize the redundancy in sequential video frames while achieving exceptional video analysis performance. 
Follow-up studies investigate extending task-oriented principle to other scenarios of mobile communications, including multi-device cooperative inference \cite{shao2022task}, image processing with digital modulation \cite{xie2023robust}, domain generalization \cite{li2023task}, and cooperative perception in autonomous driving \cite{shao2024task}.
The above studies primarily investigate conventional vision tasks such as object detection and classification, while no existing work has addressed the challenge of prohibitive transmission overhead in LMM inference. Given the surging demands for LMM inference, we adopt the task-oriented communication principle in multimodal understanding tasks. As LMMs rely on high-dimensional visual inputs and large neural networks to achieve state-of-the-art performance across diverse and complex multimodal tasks, applying the task-oriented principle to LMM inference constitutes a unique challenge.

\subsection{Neural Data Compression}
Various data compression methods have been proposed to minimize storage and transmission overhead, especially for high-dimensional visual data. 
Traditional visual data compression methods, such as JPEG \cite{wallace1991jpeg,skodras2001jpeg}, BPG \cite{BPGWeb}, and WebP \cite{webp}, adopt linear transforms \cite{goyal2001theoretical} to obtain compact representations in the frequency domain, followed by entropy coding to further remove redundancy. 
In contrast, neural compression methods \cite{Balle2017,Balle2018,Cheng2020} leverage neural networks to extract latent features and optimize the trade-off between data rate and reconstruction quality. Benefiting from the versatile nonlinear transforms \cite{NTCsurvey} and end-to-end optimization, these learning-based methods outperform traditional compression methods. Subsequent studies \cite{minnen2018joint,minnen2020channel,checkerboard,ELIC,liu2024bidirectional} equip the learnable entropy model with autoregressive structures to extract correlated contexts in both spatial and channel dimensions. These contexts exploit the intrinsic correlations in latent features to obtain a more accurate estimation of the statistics of latent features, thereby enhancing the rate-distortion (RD) performance. 
However, previous data compression methods assess reconstruction quality by traditional metrics such as peak signal-to-noise ratio (PSNR), focusing on high-fidelity image reconstruction in the pixel space. This design encourages the model to capture general features in the visual data instead of extracting specific task-relevant information, resulting in information redundancy and excessive communication overhead. 
To improve transmission efficiency, this work considers latent feature compression based on task-oriented communication principles, which eliminates the task-irrelevant redundancy in the visual features by directly optimizing the trade-off between transmission overhead and performance on downstream tasks.

\subsection{LMM Inference Acceleration}
High-resolution input images lead to a large number of visual tokens for mainstream LMMs, thus incurring prohibitive inference costs, which motivates recent studies to explore LMM inference acceleration by reducing the number of visual tokens. 
Several approaches \cite{FastV, FocusLLaVA, MustDrop} attempt to discard less important visual tokens or the key-value (KV) caches generated from these tokens during the LLM token generation process, based on the attention scores they receive.
However, as the LLM generation requires high GPU memory and extensive computation, which are only available at the edge server, visual features must be transmitted to the edge server before LLM inference.
Consequently, these methods do not alleviate the data transmission overhead in a device-edge co-inference framework.
Other works \cite{LLaVA-Mini, QueCC, TokenPacker, AVG-LLaVA, LaVIT} introduce learnable structures to select or extract critical features.
Nevertheless, these learning-based approaches require extensive training on large datasets to identify the features essential for downstream tasks, thus incurring high expenses to deploy.
Additionally, many of these approaches \cite{QueCC, TokenPacker, AVG-LLaVA, FocusLLaVA} apply identical operations to each small grid of visual features while ignoring the significant variation in semantic complexity across different regions of the image.
As a consequence, these spatially-invariant approaches fail to effectively extract a compact representation of task-relevant information, degrading the compression efficiency.
To overcome these limitations, this work proposes a training-free FMM that merges visual features before feature projection, thereby reducing both the data transmission overhead and the LLM inference latency.

\section{System Model}
\begin{figure}[!t]
    \centering
    \includegraphics[width=1\linewidth]{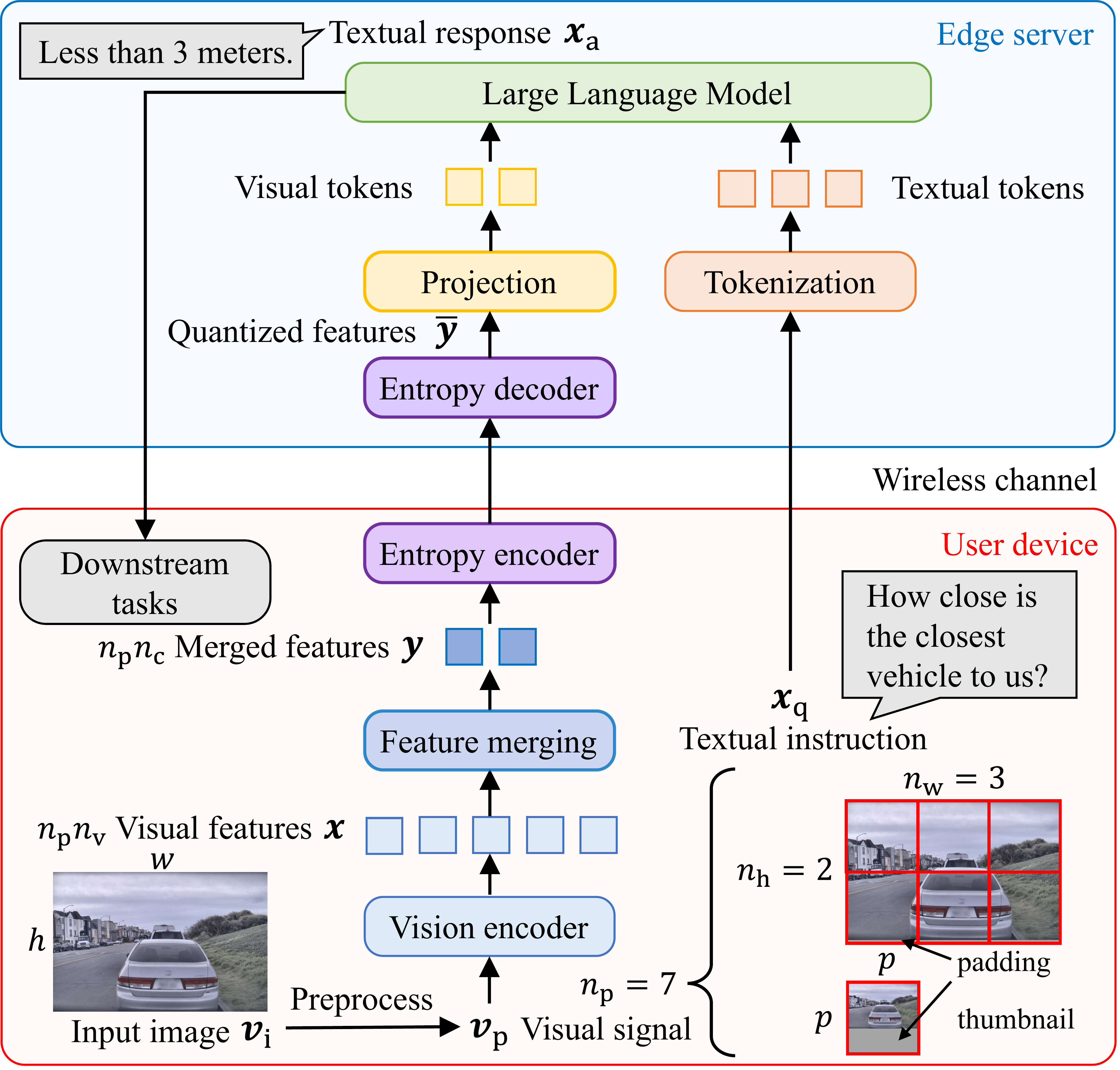}
    \caption{System diagram of the proposed feature compression method.
    The user device executes vision encoding, feature merging, and entropy encoding for the visual input, while the edge server performs entropy decoding, feature projection, and autoregressive generation of the LLM.
    The feature merging module reduces the number of the visual features to reduce both data transmission overhead and computational complexity, lowering the end-to-end system latency.
    The learnable and selective entropy model with a hyperprior is utilized to encode and decode the merged features, further reducing data transmission overhead.
    }
    \label{fig_sys}
\end{figure}
As illustrated in Fig. \ref{fig_sys}, we propose a device-edge co-inference system for LMMs, where the user device processes and encodes the input image before transmitting the encoded visual features, along with the textual instruction, to an edge server for LMM inference.
Specifically, the user device takes input of two modalities: a textual instruction that describes the user's request in natural language and a visual input that contains semantic information necessary for downstream tasks.
The textual instruction is directly transmitted to the edge server due to its small data size.
On the other hand, the input image accounts for the majority of the transmission overhead. 
Each image is represented as a tensor of pixel-wise RGB intensities $\boldsymbol{v}_{\rm i} \in \mathbb{R}^{h \times w \times 3}$, where $h$ and $w$ represent the height and width of the image, respectively.
To preserve the fine-grained details in the input image, it is enlarged, padded and partitioned into $n_{\rm h} \times n_{\rm w}$ patches \cite{llava-ov}, each with a resolution of $p \times p$ that matches the input size of the vision encoder \cite{SigLIP}.
Additionally, a thumbnail of the input image is created by padding the original image to a square shape and resizing it to the size of one patch as $p \times p$, which captures the overall structure of the image \cite{llava-ov}. 
The visual signal consists of both the detailed patches and the overall thumbnail, written as $\boldsymbol{v}_{\rm p} \in \mathbb{R}^{n_{\rm p} \times p \times p \times 3}$, where $n_{\rm p} = n_{\rm h} \times n_{\rm w} + 1$ denotes the total number of patches.

The user device leverages a pre-trained SigLIP vision encoder \cite{SigLIP} to transform the visual signal into visual features, expressed as: 
\begin{equation}
    \boldsymbol{x} = f_{\rm vision}(\boldsymbol{v}_{\rm p}),
    \label{eq_f_vision}
\end{equation}
where $f_{\rm vision}(\cdot)$ denotes the vision encoder and $\boldsymbol{x} \in \mathbb{R}^{n_{\rm p} \times n_{\rm v} \times d_{\rm v}}$ represents visual features, with $n_{\rm v}$ and $d_{\rm v}$ denoting the number of visual features and the dimension of each visual feature, respectively.
To reduce data size and accelerate inference, we propose a feature merging module that transforms the visual features $\boldsymbol{x}$ into merged features $\boldsymbol{y} \in \mathbb{R}^{n_{\rm p} \times n_{\rm c} \times d_{\rm v}}$, where $n_{\rm c}$ is the number of merged features.
These merged features are encoded into a bitstream using a learnable entropy encoder \cite{Balle2018} and then transmitted through a wireless channel to the edge server.
Following \cite{TOCvideo, shao2022task}, we consider the uplink bandwidth constraints on edge devices instead of explicitly modeling the physical channel, as channel coding and signal modulation are beyond the scope of this paper.

At the edge server, the received textual instruction is tokenized using the codebook of the pre-trained LLM.
Meanwhile, the received bitstream is decoded by an entropy decoder to recover the merged features $\hat{\boldsymbol{y}}$.
Subsequently, a two-layer multilayer perceptron (MLP) projects the merged features into the word embedding space, thus aligning the visual input with the textual tokens \cite{llava-ov}.
The resulting visual tokens are concatenated with the textual tokens, constituting the input of the LLM.
The tokens generated by the LLM, denoted by $\boldsymbol{x}_{\rm a}$, are transformed into plain text, and this textual response is transmitted back to the user device for downstream tasks.

\section{Task-Oriented Feature Compression}
We propose a task-oriented feature compression (TOFC) method for multimodal understanding in a device-edge co-inference framework, where visual features are merged by clustering and encoded by a learnable and selective entropy model before feature projection, thus reducing data transmission overhead and end-to-end system latency.
The proposed feature merging module (FMM) utilizes density peaks clustering based on $K$ nearest neighbors (DPC-KNN) \cite{DPC-KNN} to reduce the number of visual features, thereby minimizing both data transmission and computational complexity.
Subsequently, we employ a learnable entropy model with hyperprior \cite{Balle2018} to encode and decode merged features, further reducing transmission overhead.
To improve the accuracy of the probability distribution estimated by the entropy model, multiple entropy models are adaptively selected based on the characteristics of the visual features, thus enhancing compression efficiency.
The following subsections elaborate on the details of these three components.

\subsection{Clustering-Based Feature Merging}
The FMM partitions the $n_{\rm v}$ visual features in each patch into $n_{\rm c}$ clusters and averages the features in each cluster to obtain the merged features.
As a result, the number of visual features is reduced from $n_{\rm v}$ to $n_{\rm c}$, equal to only 1.1\% to 4.4\% of the original number of features $n_{\rm v}$.
This reduction significantly decreases the data size of the visual features transmitted to the edge server.
Moreover, the FMM minimizes the number of visual tokens processed by the LLM, leading to substantial reductions in GPU memory usage and computational load at the edge server.
The reductions in both data transmission and computational complexity contribute to lower end-to-end system latency, thus enabling delay-sensitive tasks and enhancing user experience.

Let $d(\boldsymbol{x}_{n,i}, \boldsymbol{x}_{n,j})$ denote the Euclidean distance between the $i$-th and $j$-th visual features in the $n$-th patch ($i,j \in \{ 1,2,\ldots,n_{\rm v} \}$, $n \in \{ 1,2,\ldots,n_{\rm p} \}$), and $\text{KNN}(\boldsymbol{x}_{n,i}, K)$ represent the set of the $K$ visual features in the $n$-th patch that are closest to the $i$-th feature $\boldsymbol{x}_{n,i}$, excluding the $i$-th feature itself.
The local density of the $i$-th feature in the $n$-th patch $\rho_{n,i}$ is calculated based on the distances between the $i$-th feature and its $K$ nearest neighbors \cite{DPC-KNN}, which is expressed as:
\begin{equation}
    \rho_{n,i} = \exp(-\frac{1}{K} \sum_{\boldsymbol{x}_{n,j} \in \text{KNN}(\boldsymbol{x}_{n,i},K)} d(\boldsymbol{x}_{n,i}, \boldsymbol{x}_{n,j})^2)
    .
    \label{eq_rho}
\end{equation}
Features with higher local densities are more adjacent to surrounding features and thus are more suitable as cluster centers.
To ensure that cluster centers are scattered across the entire latent space, we calculate the minimum distance between the $i$-th feature of the $n$-th patch and any other features in the same patch that have a higher local density \cite{DPC-KNN}, denoted as $\delta_{n,i}$, which is defined as:
\begin{equation}
    \delta_{n,i} = \begin{cases}
        {\min\limits_{j:\rho_{n,i} < \rho_{n,j}} d(\boldsymbol{x}_{n,i}, \boldsymbol{x}_{n,j}) ,} & {\text{if } \exists j, \rho_{n,i} < \rho_{n,j}} \\
        {\;\;\;\; \max\limits_j \;\;\;\; d(\boldsymbol{x}_{n,i}, \boldsymbol{x}_{n,j}) ,} & {\text{otherwise}}
    \end{cases}
    .
    \label{eq_delta}
\end{equation}
The visual feature with the highest local density is assigned the maximum distance so that it can be chosen as a cluster center.
For the $n$-th patch, we select $n_{\rm c}$ features with the highest values of $\rho_{n,i} \times \delta_{n,i}$ as cluster centers, and the remaining features are assigned to the cluster whose center is the closest.
Finally, we apply average pooling on visual features within each cluster to obtain the merged features $\boldsymbol{y}$.

We employ DPC-KNN for feature merging due to two primary reasons.
First, the proposed clustering-based approach is adaptive to the significant variation in semantic complexity across different regions of the image, unlike traditional spatially-invariant methods that apply identical operations to each small grid of visual features \cite{QueCC, TokenPacker, AVG-LLaVA, FocusLLaVA}.
In contrast, the proposed FMM dynamically partitions all visual features into clusters, with each cluster containing a varying number of features.
Consequently, the visual features are merged based on the semantic similarities between different regions of the image, resulting in higher compression efficiency.
Second, DPC-KNN is a training-free method that can be directly integrated with LMMs to effectively reduce visual features.
On the contrary, mainstream learning-based methods \cite{LLaVA-Mini, QueCC, TokenPacker, AVG-LLaVA, FocusLLaVA, LaVIT} require extensive training on large datasets to identify features essential for downstream tasks, thus incurring high deployment expenses.

\subsection{Learnable Entropy Model-Based Codec}
\begin{figure}[!t]
    \centering
    \includegraphics[width=1\linewidth]{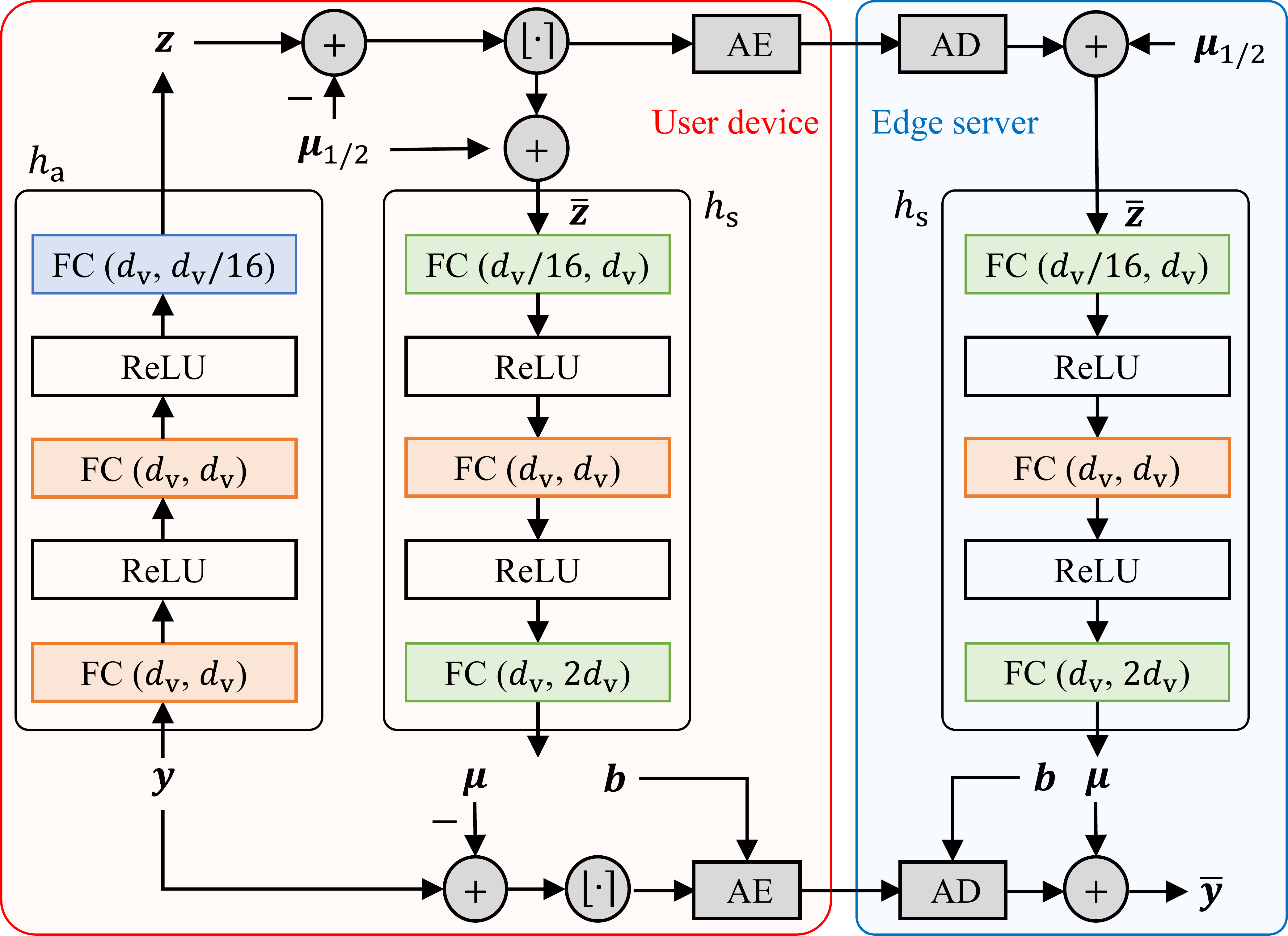}
    \caption{Network architecture of the learnable entropy model. The hyperprior analysis network $h_{\rm{a}}$ extracts the hyperprior $\boldsymbol{z}$ from the merged features $\boldsymbol{y}$. The hyperprior synthesis network $h_{\rm s}$ estimates the mean and scale of the merged features from the quantized hyperprior $\bar{\boldsymbol{z}}$. FC represents a fully connected layer, where the two parameters in parentheses denote input and output dimensions, respectively. AE and AD represent arithmetic encoder and decoder, respectively. $\lfloor \cdot \rceil$ denotes rounding to the nearest integer.}
    \label{fig_ent}
\end{figure}

\begin{figure}[!t]
    \centering
    \includegraphics[width=1\linewidth]{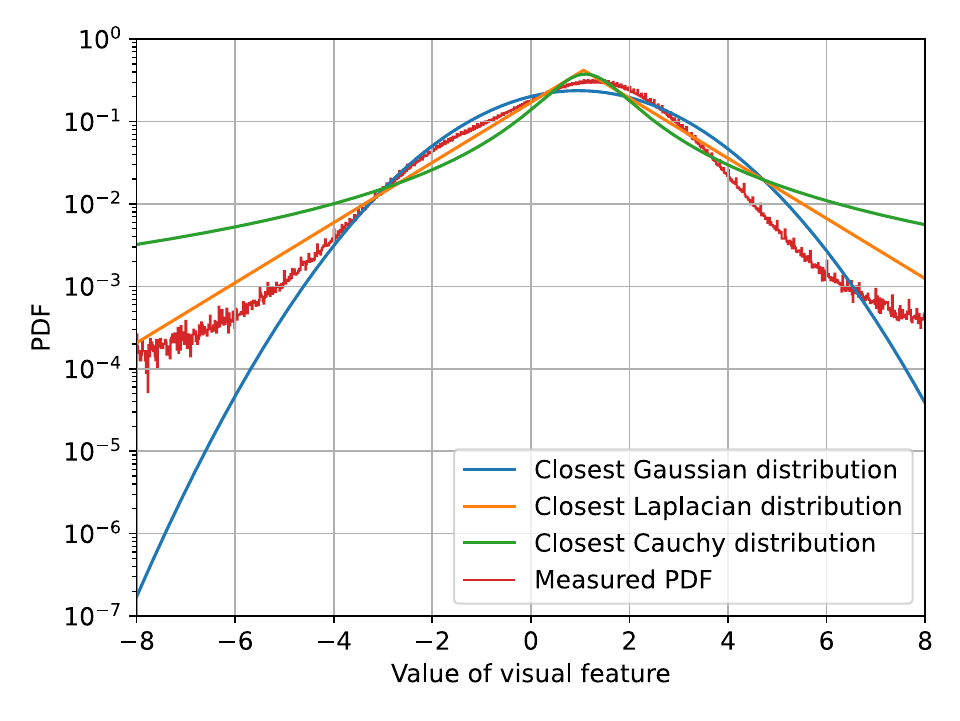}
    \caption{The measured PDF of the visual features, in comparison to Gaussian, Laplacian, and Cauchy distributions with mean and scale obtained from maximum likelihood estimation.}
    \label{fig_PDF}
\end{figure}
As the dimension of each visual feature increases significantly when transformed to the word embedding space, we employ entropy coding and transmit the bitstream before feature projection.
Effective and compact entropy coding relies on the accurate probability distribution of the discrete variable \cite{NTCsurvey}. However, it is practically infeasible to directly estimate arbitrary probability distribution. As a solution, we model the merged features $\boldsymbol{y}$ by Laplacian distribution, where the mean and scale are derived from the hyperprior $\boldsymbol{z}$.
The probability distribution of the hyperprior $\boldsymbol{z}$ is estimated by a learnable factorized entropy model \cite{Balle2018}.
Both $\boldsymbol{y}$ and $\boldsymbol{z}$ are quantized and encoded by an entropy coder based on the estimated probability distribution, thereby obtaining a more compact bitstream representation while maintaining acceptable distortion for downstream tasks.

As illustrated in Fig. \ref{fig_ent}, a hyperprior analysis network $h_{\rm a}$ is employed to extract the hyperprior $\boldsymbol{z} \in \mathbb{R}^{n_{\rm p} \times n_{\rm c} \times \frac{d_{\rm v}}{16}}$ from the merged features $\boldsymbol{y}$.
Since the entropy coder operates on integers, both $\boldsymbol{y}$ and $\boldsymbol{z}$ are quantized before entropy coding \cite{NTCsurvey}.
The mean and scale of the merged features are estimated from the quantized hyperprior $\bar{\boldsymbol{z}}$ using a hyperprior synthesis network $h_{\rm s}$ as:
\begin{equation}
    (\boldsymbol{\mu}, \boldsymbol{b}) = h_{\rm s}(\bar{\boldsymbol{z}})
    ,
    \label{eq_h_s}
\end{equation}
where $\boldsymbol{\mu}$ and $\boldsymbol{b}$ denote the estimated mean and scale of the merged features $\boldsymbol{y}$, respectively.
The quantization of merged features introduces an additive uniform noise \cite{Balle2017} as $\bar{\boldsymbol{y}} = \boldsymbol{y} + \boldsymbol{o}$,
where $\bar{\boldsymbol{y}}$ denotes the quantized merged features, $\boldsymbol{o} \sim \mathcal{U}(-0.5, 0.5)$ is the additive noise, and $\mathcal{U}(a, b)$ represents a uniform distribution from $a$ to $b$.
As illustrated in Fig. \ref{fig_PDF}, the empirical measurement of the probability density function (PDF) of the visual features resembles the Laplacian distribution.
Consequently, the probability mass function (PMF) of $\bar{\boldsymbol{y}}$ is modeled as Laplacian convolved with a uniform distribution, expressed as:
\begin{equation}
    p(\bar{\boldsymbol{y}} | \bar{\boldsymbol{z}}) = \left( \mathcal{L}(\boldsymbol{\mu}, \boldsymbol{b}) * \mathcal{U}(-0.5, 0.5) \right) (\bar{\boldsymbol{y}})
    ,
    \label{eq_p_y}
\end{equation}
where $\mathcal{L}(\boldsymbol{\mu}, \boldsymbol{b})$ represents a Laplacian distribution with a mean vector of $\boldsymbol{\mu}$ and a scale vector of $\boldsymbol{b}$, and $*$ denotes convolution.
Similarly, the PMF of quantized hyperprior $\bar{\boldsymbol{z}}$ is written as:
\begin{equation}
    p(\bar{\boldsymbol{z}}) = \left( p(\boldsymbol{z}) * \mathcal{U}(-0.5, 0.5) \right) (\bar{\boldsymbol{z}})
    ,
    \label{eq_p_z}
\end{equation}
where $p(\boldsymbol{z})$ is the estimated probability distribution of $\boldsymbol{z}$ \cite{Balle2018}.
Similar to prior works on vector quantized-variational autoencoder (VQ-VAE) \cite{vqvae} and neural image compression \cite{minnen2020channel}, we employ the straight-through estimator (STE) during training to allow the back-propagation of gradients through the non-differentiable quantization operation, expressed as:
\begin{equation}
    \bar{\boldsymbol{y}} = \rm{STE}(\boldsymbol{y} - \boldsymbol{\mu}) + \boldsymbol{\mu}
    ,
    \label{eq_bar_y}
\end{equation}
\begin{equation}
    \bar{\boldsymbol{z}} = \rm{STE}(\boldsymbol{z} - \boldsymbol{\mu}_{\rm{1/2}}) + \boldsymbol{\mu}_{\rm{1/2}} 
    ,
    \label{eq_bar_z}
\end{equation}
where ${\rm STE}(x) = {\rm sg}({\rm round}(x) - x) + x$ is the STE operation, ${\rm sg}(\cdot)$ is the stop-gradient operator, ${\rm round}(\cdot)$ denotes rounding to the nearest integer, and $\boldsymbol{\mu}_{\rm{1/2}}$ denotes the median vector of $\boldsymbol{z}$, obtained from the learned probability model \cite{Balle2018}.

During training, the entropy model learns to strike a balance between data compression and performance on downstream tasks.
Thus, the total loss $L$ consists of both the rate loss $R$ and the distortion loss $D$, written as:
\begin{equation}
    L = D + \lambda R
    ,
    \label{eq_L}
\end{equation}
where $\lambda$ is the Lagrange multiplier controlling the trade-off between rate and distortion.
The distortion loss is evaluated by the cross-entropy loss of the next token prediction as:
\begin{equation}
    D =  -\frac{1}{T} \sum_{t=1}^{T} \log p(\boldsymbol{x}_{{\rm a},t} | \boldsymbol{x}_{{\rm a}, 1}, \ldots, \boldsymbol{x}_{{\rm a},t-1})
    ,
    \label{eq_D}
\end{equation}
where $\boldsymbol{x}_{{\rm a},t}$ denotes the $t$-th token predicted by the LLM, and $T$ is the length of the ground-truth response.
The rate loss should be the joint entropy of quantized merged features and the hyperprior as
\begin{align}
    H(\bar{\boldsymbol{y}}, \bar{\boldsymbol{z}}) 
    & = H(\bar{\boldsymbol{y}} | \bar{\boldsymbol{z}}) + H(\bar{\boldsymbol{z}}) \nonumber \\ 
    & = \mathbb{E}_{\boldsymbol{v}_{\rm i} \sim p_{\boldsymbol{v}_{\rm i}}} [-\log p_{\bar{\boldsymbol{y}} | \bar{\boldsymbol{z}}} (\bar{\boldsymbol{y}} | \bar{\boldsymbol{z}}, \boldsymbol{v}_{\rm i}) - \log p_{\bar{\boldsymbol{z}}} (\bar{\boldsymbol{z}} | \boldsymbol{v}_{\rm i})]
    ,
    \label{eq_joint_entropy}
\end{align}
where $H(\cdot)$ denotes the Shannon entropy, $\mathbb{E}[\cdot]$ is the expectation operator, and $p_{\boldsymbol{v}_{\rm i}}$ represents the distribution of the input image.
However, the true posterior of the visual features given the input image is assumed intractable \cite{Balle2018}, and the joint entropy can not be directly derived from (\ref{eq_joint_entropy}).
To address this issue, a tractable upper bound of the joint entropy is formulated based on the conditional entropy model $p(\bar{\boldsymbol{y}} | \bar{\boldsymbol{z}})$ and the factorized entropy model $p(\bar{\boldsymbol{z}})$ \cite{TOCvideo}, expressed as:
\begin{equation}
    H(\bar{\boldsymbol{y}}, \bar{\boldsymbol{z}}) \le \mathbb{E}_{\bar{\boldsymbol{y}} \sim p(\bar{\boldsymbol{y}} | \bar{\boldsymbol{z}})} [-\log p(\bar{\boldsymbol{y}} | \bar{\boldsymbol{z}}) ] + \mathbb{E}_{\bar{\boldsymbol{z}} \sim p(\bar{\boldsymbol{z}})} [-\log p(\bar{\boldsymbol{z}})]
    .
    \label{eq_entropy_bound}
\end{equation}
We derive the rate loss from this upper bound as:
\begin{equation}
    R = -\frac{1}{n_{\rm p}n_{\rm c}d_{\rm v}} (\log p(\bar{\boldsymbol{y}} | \bar{\boldsymbol{z}}) + \log p(\bar{\boldsymbol{z}}))
    .
    \label{eq_R}
\end{equation}
Apart from training the parameters of the entropy model, we employ low-rank adaptation (LoRA) \cite{LoRA} on the vision encoder to adjust the probability distribution of the visual features, allowing a more flexible data size.
We also train the parameters of the MLP projector to accommodate the modifications in visual features introduced by feature merging and quantization.
Additionally, the LLM is fine-tuned using LoRA to bridge the gap caused by the compaction of visual features.

\subsection{Selective Entropy Model}
\begin{figure*}[!t]
    \centering
    \includegraphics[width=.85\linewidth]{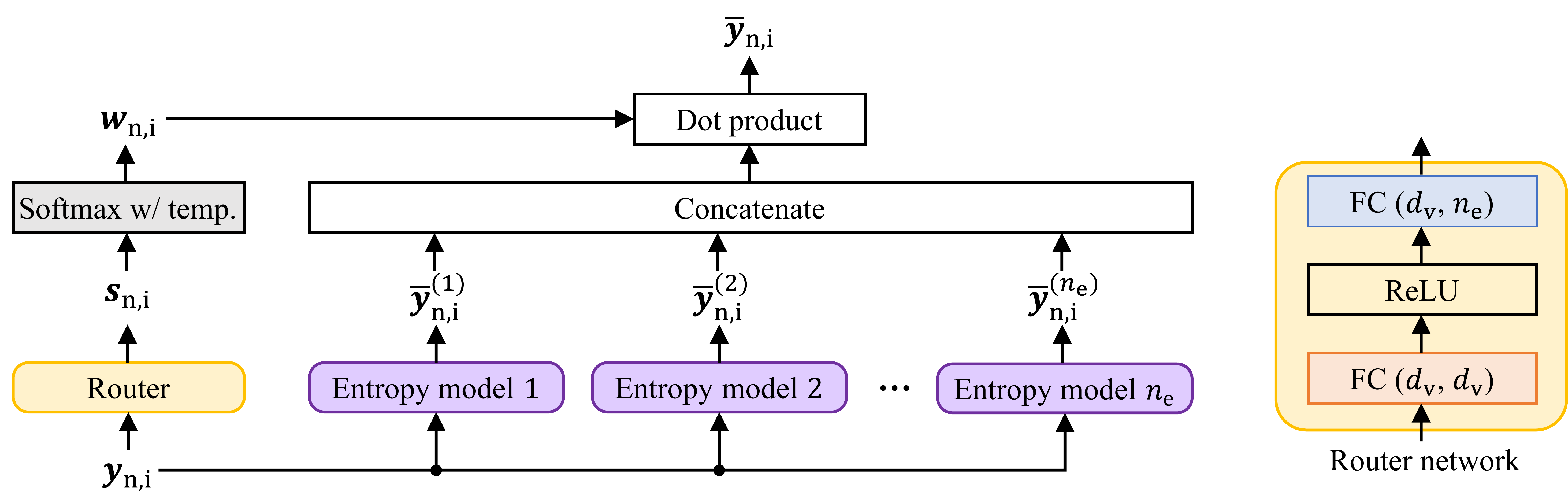}
    \caption{Network architecture of the selective entropy model (w/ temp.: with temperature). 
    The router network analyzes each merged feature $\boldsymbol{y}_{n,i}$ and generates a score vector $\boldsymbol{s}_{n,i}$, which is processed by a softmax function with temperature $T$ to compute the weighting coefficients for $n_{\rm e}$ entropy models.
    During training, the final quantized features $\bar{\boldsymbol{y}}_{n,i}$ are the weighted sum of the decoding results from each entropy model.
    During inference, with $T \rightarrow 0$, only the entropy model with the highest score is used for encoding and decoding.
    }
    \label{fig_MoE}
\end{figure*}
Considering the semantic diversity of images, a single entropy model cannot accurately estimate the statistics of the merged visual features.
An inaccurate probability model introduces overhead during entropy coding, resulting in a bitstream length that exceeds the actual entropy of the quantized features.
To eliminate this redundancy, we employ multiple entropy models that can be adaptively selected based on the characteristics of the visual features.
Each entropy model is specialized in encoding specific features, thus achieving a data rate closer to the actual entropy and improving the overall compression efficiency.

As illustrated in Fig. \ref{fig_MoE}, we utilize a router network to analyze the characteristics of merged features and generate a score for each entropy model.
Specifically, the score vector for the $i$-th feature in the $n$-th patch is expressed as
\begin{equation}
    \boldsymbol{s}_{n,i} = f_{\rm router}(\boldsymbol{y}_{n,i})
    ,
    \label{eq_f_router}
\end{equation}
where the length of $\boldsymbol{s}_{n,i}$ is equal to the number of entropy models $n_{\rm e}$.
The idea of adaptive routing and activation of model weights is inspired by the mixture-of-experts (MoE) architecture widely adopted in LLMs \cite{SwitchTransformer, deepseekv3}, which remains unexplored for learnable entropy models.
Based on the unique characteristics of entropy models, we employ different strategies in the training and inference processes.
During training, all entropy models encode and decode merged features separately, and the final quantized features are the weighted sum of the decoding results from each entropy model, given by:
\begin{equation}
    \bar{\boldsymbol{y}}_{n,i} = \sum_{e=1}^{n_{\rm e}} w_{n,i}^{(e)} \bar{\boldsymbol{y}}_{n,i}^{(e)}
    ,
    \label{eq_MoE_bar_y}
\end{equation}
where $w_{n,i}^{(e)}$ and $\bar{\boldsymbol{y}}_{n,i}^{(e)}$ represent the weighting coefficient and the decoding result of the $i$-th feature in the $n$-th patch using the $e$-th entropy model.
Similarly, the rate loss in (\ref{eq_R}) is substituted by the weighted sum of the estimated entropies from all entropy models as:
\begin{equation}
    R = -\frac{1}{n_{\rm p}n_{\rm c}d_{\rm v}} \sum_{e=1}^{n_{\rm e}} w_{n,i}^{(e)} (\log p(\bar{\boldsymbol{y}}^{(e)} | \bar{\boldsymbol{z}}^{(e)}) + \log p(\bar{\boldsymbol{z}}^{(e)}))
    .
    \label{eq_MoE_R}
\end{equation}
The weighting coefficient for each entropy model is calculated from the score generated by the router network as:
\begin{equation}
    w_{n,i}^{(e)} = \frac{\exp (s_{n,i}^{(e)}/T)} {\sum_{e=1}^{n_{\rm e}} \exp (s_{n,i}^{(e)}/T)}
    ,
    \label{eq_w}
\end{equation}
where $T$ is the temperature that controls the distribution of the weighting coefficients.
The temperature is high at the beginning of training, and the weighting coefficients of different entropy models are similar, allowing all entropy models to learn the basic coding capabilities.
The temperature gradually decreases as training progresses, and the weighting coefficients converge towards the single optimal selection employed during inference, which enables the specialization of different entropy models.
To avoid the scenario where certain entropy models are never selected, we introduce balance loss to the RD loss in (\ref{eq_L}), expressed as:
\begin{equation}
    L_{\rm balance} = \alpha \sum_{e=1}^{n_{\rm e}} (\frac{1}{n_{\rm p} n_{\rm c}} \sum_{n=1}^{n_{\rm p}} \sum_{i=1}^{n_{\rm c}} w_{n,i}^{(e)} - \frac{1}{n_{\rm e}})^2
    ,
    \label{eq_L_balance}
\end{equation}
where $\alpha$ is the weight of the balance loss.
A high value of $\alpha$ encourages a uniform usage frequency across entropy models in each batch but hinders the specialization of each entropy model.
Thus, the value of $\alpha$ is carefully selected via experiments to maximize the compression efficiency.
The detailed training procedures are summarized in Algorithm \ref{alg_train}.
\begin{algorithm*}[t]
    \caption{Training Procedures of the Proposed TOFC Method}
    \label{alg_train}
    \begin{algorithmic}[1]
    \Require Training dataset, initialized model parameters, hyper-parameters $\lambda$, $T$, $\alpha$.
    \Ensure Model parameters.
    \While{batch $b = 1$ to $B$}
        \State Process the input image $\boldsymbol{v}_{\rm i}$ into multiple patches $\boldsymbol{v}_{\rm p}$. Extract visual features $\boldsymbol{x}$ using the vision encoder as in (\ref{eq_f_vision}).
        \State Calculate the local density $\rho$ and the distance $\delta$ of visual features as in (\ref{eq_rho}) and (\ref{eq_delta}), respectively.
        \State Select the $n_{\rm c}$ visual features with the largest $\rho \times \delta$ as cluster centers for each image patch.
        \State Assign remaining features to the closest cluster center. Apply average pooling on all clusters to obtain $\boldsymbol{y}$.
        \While{$e = 1$ to $n_{\rm e}$}
            \State Extract the hyperprior $\boldsymbol{z}$ from $\boldsymbol{y}$. Use STE to quantize the hyperprior as in (\ref{eq_bar_z}). 
            \State Calculate statistics of $\boldsymbol{y}$ using quantized hyperprior $\bar{\boldsymbol{z}}$ as in (\ref{eq_h_s}). Use STE to quantize merged features as in (\ref{eq_bar_y}).
        \EndWhile
        \State Analyze $\boldsymbol{y}$ using router network as in (\ref{eq_f_router}). Calculate the weighting coefficients of all entropy models as in (\ref{eq_w}).
        \State Calculate final features as in (\ref{eq_MoE_bar_y}), followed by feature projection to obtain visual tokens. Conduct LLM inference.
        \State Calculate loss function $L = D + \lambda R + L_{\rm balance}$, where $D$, $R$ and $L_{\rm balance}$ are given in (\ref{eq_D}), (\ref{eq_MoE_R}) and (\ref{eq_L_balance}), respectively.
        \State Update model parameters through backpropagation. Update the temperature $T$.
    \EndWhile
    \end{algorithmic}
\end{algorithm*}

During inference, each merged feature is encoded and decoded using the entropy model with the highest score to minimize data transmission, which is equivalent to temperature approaching zero.
The weighting coefficient is expressed as:
\begin{equation}
    w_{n,i}^{(e)} = \begin{cases}
        {1,} & {\text{if } \forall e' \neq e, \boldsymbol{s}_{n,i}^{(e)} > \boldsymbol{s}_{n,i}^{(e')}} \\
        {0,} & {\text{otherwise}}
    \end{cases}
    .
\end{equation}
The indices of the selected entropy model are transmitted to the edge server as side information, consuming a negligible data size compared to that of encoded features.

\section{Experiment Results}
\subsection{Experiment Setup} \label{sec:exp_setup}
\subsubsection{Network Implementation}
We employ LLaVA-OneVision-7B \cite{llava-ov} as the backbone model and SigLIP-SO400M \cite{SigLIP} as the vision encoder.
The resolution of each image patch is $p \times p = 384 \times 384$ pixels, and the number of patches in the height and width dimensions are given by $n_{\rm h} = \lceil \frac{h}{p} \rceil$ and $n_{\rm w} = \lceil \frac{w}{p} \rceil$, respectively, where $\lceil \cdot \rceil$ denotes rounding up to the nearest integer.
The number of visual features for each patch is $n_{\rm v} = 729$, and the dimension of each visual feature is $d_{\rm v} = 1152$.
The number of merged features $n_{\rm c}$ varies from 8 to 32, and the number of entropy models is $n_{\rm e} = 16$.
The number of nearest neighbors is fixed at $K = 5$, as we find negligible variations in task performance after adjusting its value.
The learnable entropy models are implemented based on the CompressAI package \cite{compressai}.
The structures of each learnable entropy model and the router network are illustrated in Figs. \ref{fig_ent} and \ref{fig_MoE}, respectively.

\subsubsection{Training Details}
The training data consists of the 665K instruction tuning samples utilized in the second training stage of LLaVA-1.5 \cite{llava1.5}, accounting for a small portion of 4.8M samples used in backbone model training.
The proposed model is trained for only one epoch, with the trainer settings identical to those used in the second training stage of the backbone model \cite{llava-ov}.
With the rank of LoRA adapters \cite{LoRA} set to 64, the total number of trainable parameters is approximately 300 million, only a small percentage of the total parameter count of 8 billion.
The Lagrange multiplier $\lambda$ varies between 0.003 and 0.03 to achieve different trade-offs between data transmission overhead and the performance of downstream tasks.
The weight of the balance loss is set as $\alpha = 0.001$.
Additionally, the temperature of softmax function $T$ exponentially decreases from 10 to 0.1 throughout the training process.
The training process takes 8.1 hours on a server with eight RTX 4090 GPUs, substantially lower than the backbone model \cite{llava-ov} which is trained for 66 hours in total on 256 A100 GPUs or 128 H100 GPUs.

\subsubsection{Evaluation Details}
The image understanding capability is measured by the following seven VQA benchmarks.
\begin{itemize}
    \item The RealWorldQA benchmark evaluates real-world spatial understanding via 765 multiple-choice questions on high-resolution images, which require scene detail recognition and commonsense reasoning.
    \item The MME benchmark \cite{MME} employs custom-designed, manually annotated instruction-answer pairs to measure both perception and cognition abilities of LMMs across 14 distinct subtasks.
    \item The test set of AI2 diagrams (AI2D) \cite{AI2D} is comprised of 3088 grade-school science diagrams with question-answer pairs to evaluate the diagram understanding capabilities.
    \item The English development set of MMBench \cite{MMB} comprises 4329 multiple-choice questions that assess diverse multimodal understanding capabilities of LMMs.
    \item The MMStar benchmark \cite{MMStar} collects 1500 human-selected and validated samples, evaluating 6 core capabilities and 18 detailed axes of LMMs.
    \item The test set of ScienceQA \cite{ScienceQA} includes 2017 multiple choice questions in natural science, social science, and linguistics, sourced from elementary and high school science curricula.
    \item The validation set of massive multi-discipline multimodal understanding (MMMU) benchmark \cite{MMMU} contains 1050 multimodal questions from college exams, quizzes, and textbooks spanning 30 subjects to evaluate expert-level reasoning capabilities. 
\end{itemize}
\noindent The implementation of evaluation is based on the VLMEvalKit package \cite{vlmevalkit}.
The user device utilized for testing is an NVIDIA Jetson AGX Orin, and the edge server is equipped with an adequate number of NVIDIA RTX 4090 GPUs.
We compare the proposed method against four data-oriented image compression methods, including traditional methods JPEG \cite{skodras2001jpeg} and WebP \cite{webp}, and neural codecs hyperprior \cite{Balle2018} and ELIC \cite{ELIC}.
To ensure a fair comparison in terms of system latency, we train a model with the proposed FMM using the distortion loss in (\ref{eq_D}).
This model is utilized when measuring the rate-performance curves of the four data-oriented image compression methods.

\subsection{Results Analysis}
\subsubsection{Rate-Performance Curve}
\begin{figure}[!t]
    \centering
    \subfloat[]{\includegraphics[width=1\linewidth]{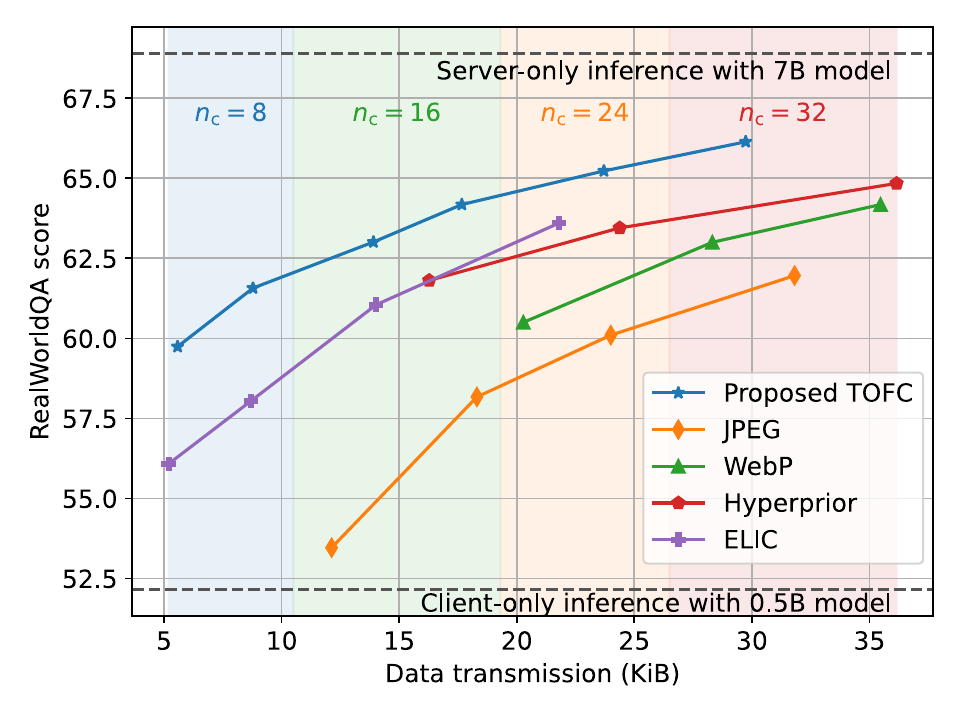}} \hfill
    \subfloat[]{\includegraphics[width=1\linewidth]{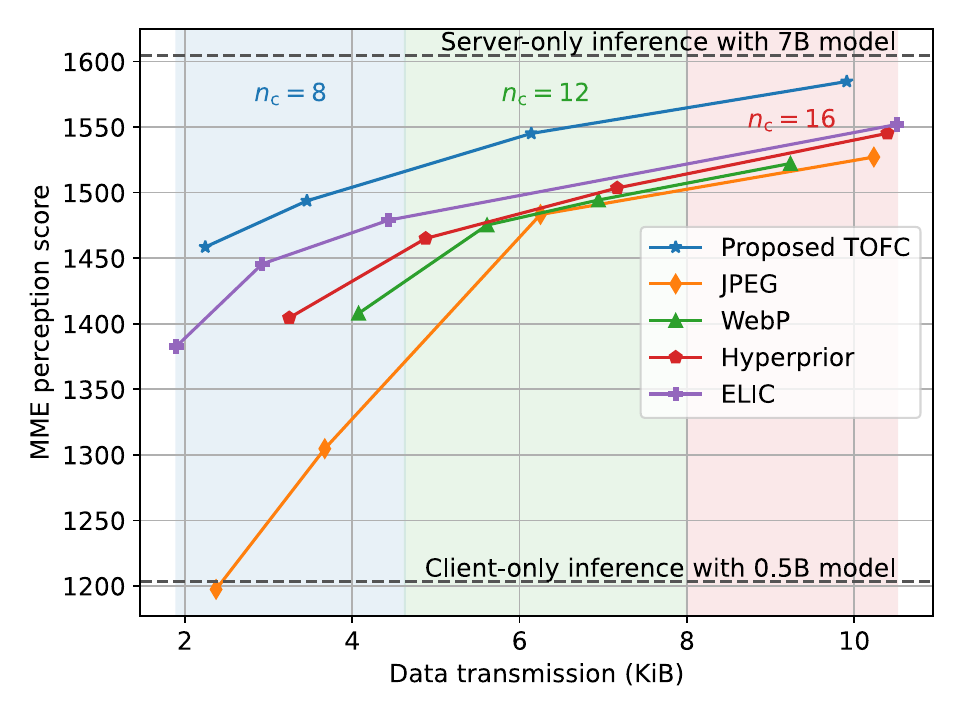}} \hfill
    \caption{The rate-performance curves of different methods in VQA benchmarks. Subfigures (a) and (b) show results on RealWorldQA and MME benchmarks, respectively.}
    \label{fig_RD_MME_Real}
\end{figure}
\begin{figure}[!t]
    \centering
    \includegraphics[width=1\linewidth]{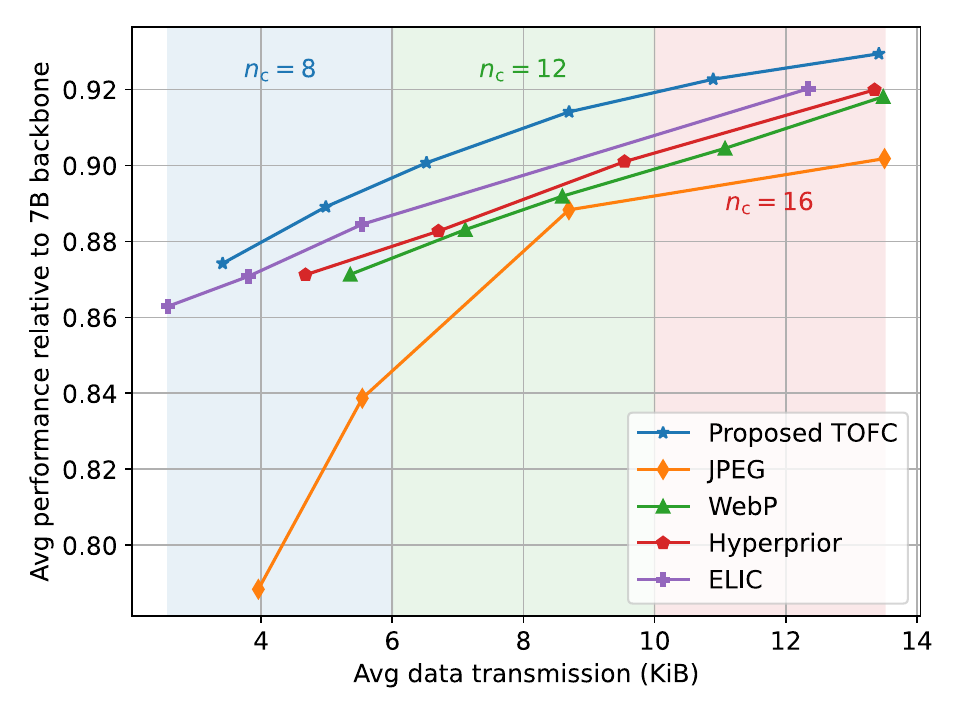}
    \caption{The rate-performance curves of different methods on the seven VQA benchmarks elaborated in Subsection V-A-3. The task performance is assessed based on the average performance across all benchmarks, normalized by the performance of the 7B backbone.}
    \label{fig_RD_avg}
\end{figure}
We train multiple instances of proposed TOFC models with varying Lagrange multipliers $\lambda$ and numbers of merged features $n_{\rm c}$ to explore the trade-off between data transmission overhead and task performance.
Note that the actual data size of encoded symbols is slightly higher than the entropy derived from estimated probabilities due to the integer symbol length inherent to arithmetic entropy coding, and we report the former value to allow a fair comparison.
To traverse on the rate-performance curves, we adjust quality settings for traditional JPEG \cite{skodras2001jpeg} and WebP \cite{webp} compression methods, and change the Lagrange multiplier $\lambda$ for neural codecs hyperprior \cite{Balle2018} and ELIC \cite{ELIC}.
Additionally, we integrate these data-oriented compression methods with the FMM proposed in Section IV-A to ensure an LLM inference latency identical to TOFC at the edge server.
Fig. \ref{fig_RD_MME_Real} presents the rate-performance curves of different methods on RealWorldQA and MME benchmarks.
For the RealWorldQA benchmark with high-resolution images, the proposed TOFC method achieves 30\% to 52\% reductions in data transmission across various target task performances in comparison to ELIC, despite lower encoding complexity.
The reductions are even more pronounced when compared to the hyperprior codec, consistently exceeding 40\%.
Traditional WebP and JPEG compressions struggle to maintain comparable task performance, incurring up to 2.89$\times$ and 3.84$\times$ data transmission overhead, respectively.
For the MME benchmark with low-resolution images, the proposed TOFC method still provides approximately 36\% and 46\% reductions in data transmission, compared to ELIC and hyperprior codecs, respectively.
Traditional WebP and JPEG compressions require 2.33$\times$ and 2.65$\times$ data transmission, respectively, to achieve the same task performance as the TOFC method.
These results demonstrate the effectiveness of the TOFC method in discarding redundant information in visual features while maintaining task performance.

Fig. \ref{fig_RD_avg} demonstrates the rate-performance curves of different methods on the seven VQA benchmarks elaborated in Subsection V-A-3.
The task performance is assessed based on the average performance across all benchmarks, normalized by the performance of the 7B backbone.
The proposed TOFC method achieves approximately 23\% and 32\% reductions in data transmission, compared with ELIC and hyperprior codecs, respectively.
In comparison to traditional JPEG and WebP compressions, TOFC provides even more substantial data size savings of 47\% and 37\%, respectively.
These results underscore the efficiency of the proposed TOFC method across a variety of VQA benchmarks.
Our TOFC method minimizes data transmission overhead while maintaining high task performance, thereby enhancing user experience and enabling timely response for delay-sensitive tasks.

\subsubsection{Latency Analysis}
\begin{figure}[!t]
    \centering
    \subfloat[]{\includegraphics[width=1\linewidth]{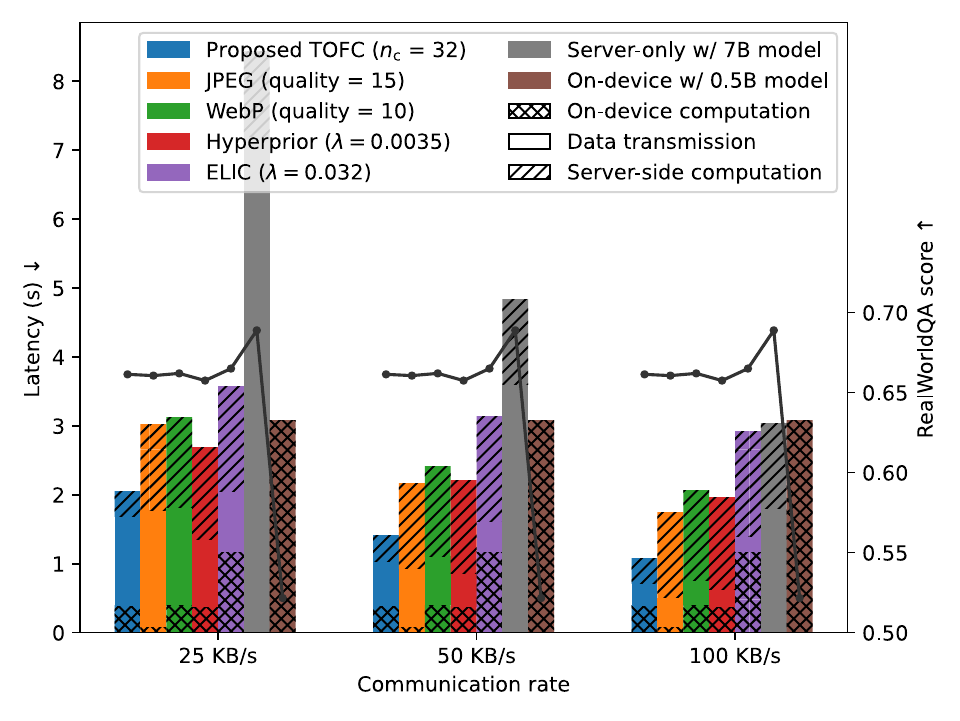}} \hfill
    \subfloat[]{\includegraphics[width=1\linewidth]{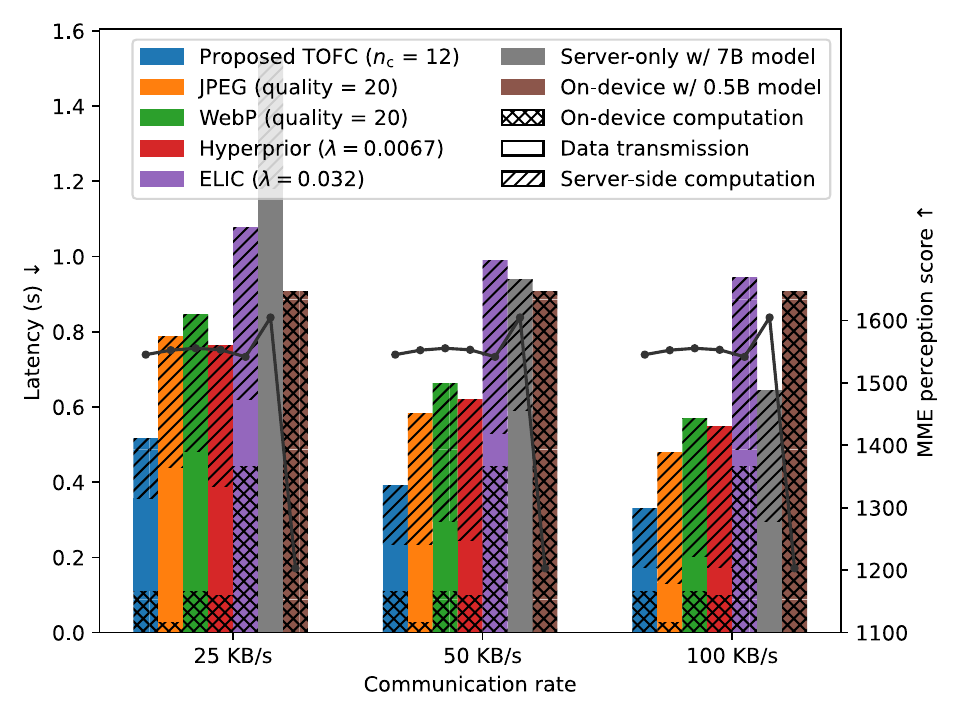}}
    \caption{Average inference latency per user request (the bar plot corresponding to the left y-axis) and benchmark score (the line plot corresponding to the right y-axis) under different communication rates. The total latency consists of on-device computation time, data transmission latency, and server-side computation time. Subfigures (a) and (b) show results on RealWorldQA and MME benchmarks, respectively.}
    \label{fig_latency}
\end{figure}
To evaluate the low latency advantage of our TOFC, we conduct a comparison of the end-to-end system latency across different methods, where the total latency consists of on-device computation time, data transmission latency, and server-side computation time.
To ensure a fair comparison, the Lagrange multiplier $\lambda$ for neural codecs and the quality settings for traditional methods are carefully selected to achieve similar task performance.
Fig. \ref{fig_latency} demonstrates the average inference latency per user request and the benchmark score under different communication rates on the RealWorldQA and MME benchmarks.

For the RealWorldQA benchmark with high-resolution images, the proposed TOFC method achieves system latency reductions between 32\% and 47\% compared with traditional image compression methods, JPEG and WebP, under various communication rates.
The system latency of WebP compression is slightly higher than that of JPEG compression, as the increased on-device computation time outweighs the reduced data transmission latency.
Compared with ELIC, TOFC achieves even more significant latency reductions from 43\% to 63\%, since the complex context calculation in modern neural compression methods incurs excessive on-device computation time.
At higher communication rates, the latency reduction becomes more prominent, as TOFC significantly accelerates LLM inference at the edge server by reducing the number of visual tokens.
These results demonstrate the superior latency performance of TOFC compared to data-oriented image compressions, benefiting from the proposed FMM method, which significantly reduces computation time at the edge server.
As for the traditional single-node computation, both server-only and device-only inference schemes suffer from excessive latencies of over 3 seconds, causing severe degradations on the user experience.
Moreover, even at high communication rates, the average data transmission latency is 1.8 seconds without compression, constituting the majority of system latency in mainstream server-only inference schemes.
These results highlight the necessity of minimizing data transmission overhead,  especially when the resolution of the input image is high.

For the MME benchmark with low-resolution images, the proposed TOFC method manages to complete inference in less than half the time compared to ELIC.
Compared with JPEG and WebP compressions, the inference latency of TOFC are approximately 67\% and 59\% those of the JPEG and WebP compressions.
The system latency of server-only inference ranges between 0.64 seconds and 1.53 seconds, lower than that observed in the RealWorldQA benchmark due to the lower image resolution.
However, the TOFC method still reduces inference time to only one-third of that required by server-only inference under low communication rate, without noticeable degradation in task performance.
As for device-edge co-inference where the uncompressed visual features are directly transmitted to the edge server, the benchmark score is the same as server-only inference with 7B model, but the latency is 40 to 60 times higher, thus not illustrated in Fig. \ref{fig_latency}.

\subsubsection{Case Study}
\begin{figure}[!t]
    \centering
    \subfloat[]{\includegraphics[width=1\linewidth]{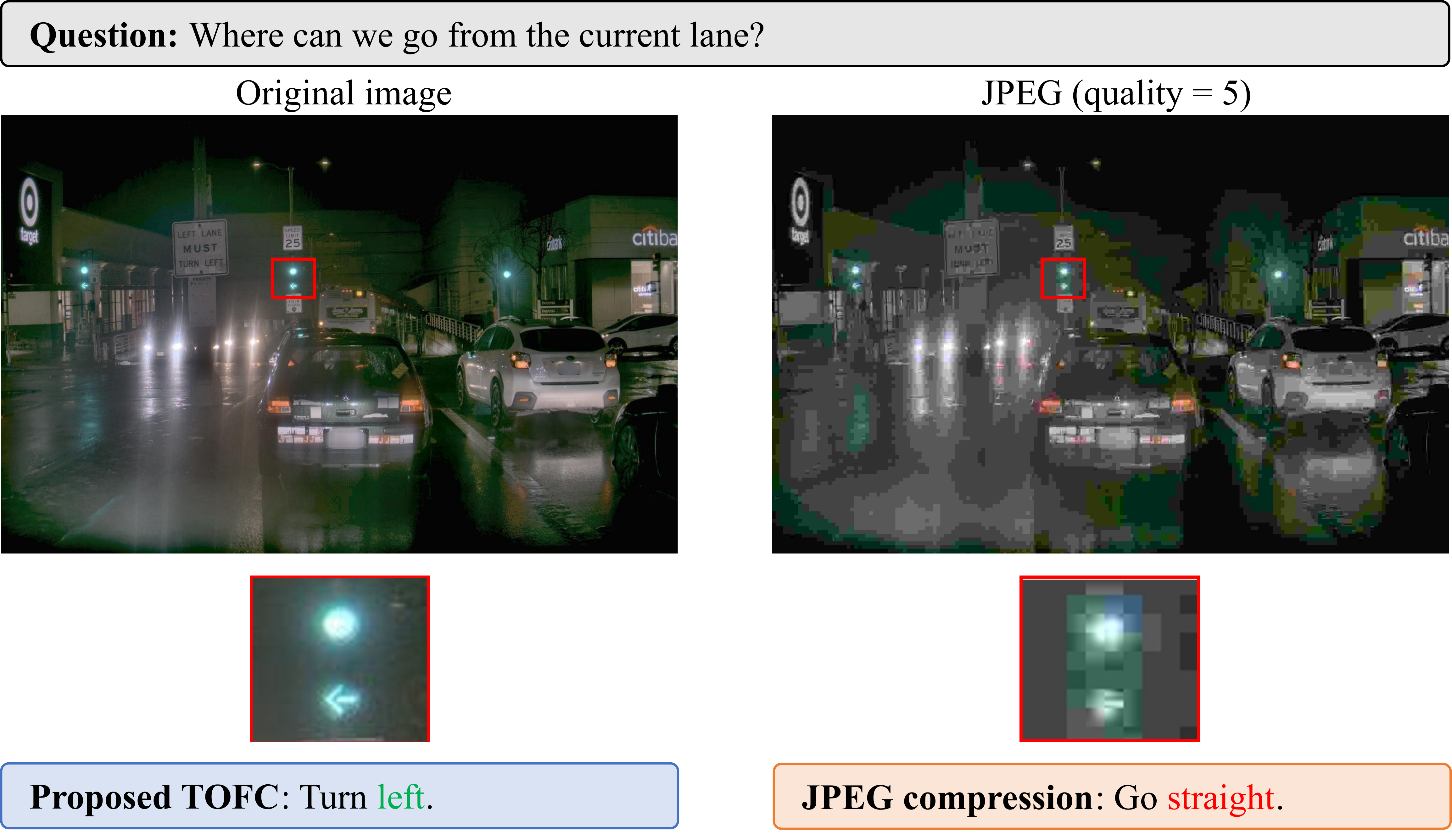}} \hfill
    \subfloat[]{\includegraphics[width=1\linewidth]{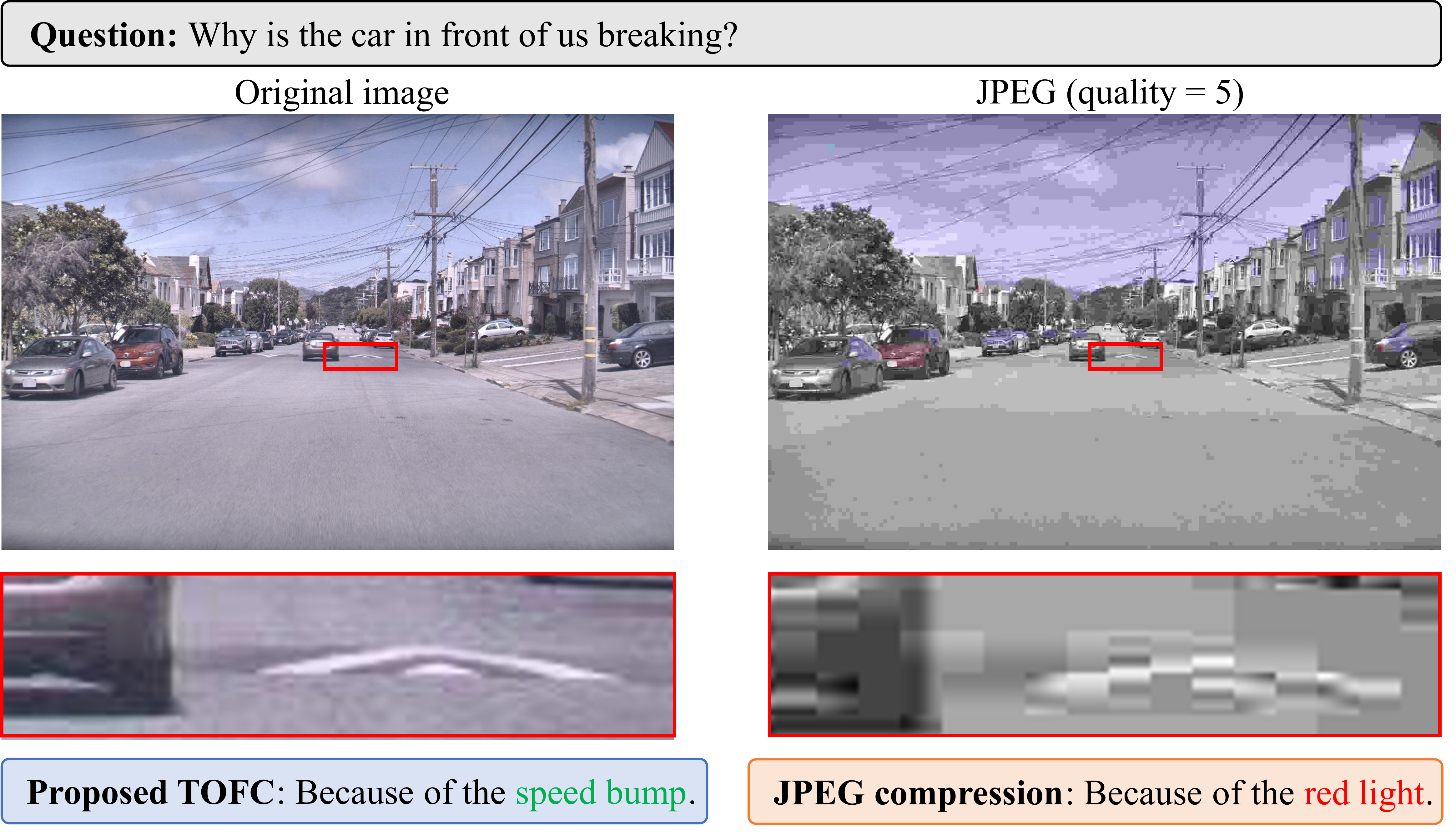}}
    \caption{Qualitative comparisons between the proposed TOFC method and traditional JPEG compression. The TOFC method performs feature merging and encoding in the latent space to retain critical information in the image, thus responding with the correct answer. In contrast, JPEG compression distorts the image at the pixel level, leading to severe degradation of image semantics. Both examples in (a) and (b) are from the RealWorldQA benchmark.}
    \label{fig_case_study}
\end{figure}
To intuitively demonstrate the advantages of the proposed TOFC method, we present examples from the RealWorldQA benchmark, where traditional JPEG compression struggles to preserve image semantics.
Fig. \ref{fig_case_study} shows two qualitative comparisons of the TOFC method and traditional JPEG compression.
The proposed TOFC method performs feature merging and encoding in the latent space to retain critical information in the image, thus responding with the correct answer.
In contrast, JPEG compression distorts the image at the pixel level, leading to significant degradation of image semantics that adversely affects downstream visual tasks.
In Fig. \ref{fig_case_study}(a), the left arrow on the green traffic light is distorted beyond recognition.
As a result, the LLM misinterprets the shape of the light as a solid circle, leading to an erroneous inference of driving straight rather than a left turn.
In Fig. \ref{fig_case_study}(b), the speed bump on the road becomes indistinguishable due to severe pixelation caused by JPEG compression.
Consequently, the LLM incorrectly infers that the car stops because of a red traffic light, which is absent in the image.
These examples evidence the limitations of traditional pixel-based image compression methods, which severely impair task-related information under low data size constraints.
Conversely, the proposed TOFC method is trained to preserve critical visual semantics, thereby enhancing task performance while maintaining low data transmission overhead.

\subsubsection{Extension to other LMMs}
To demonstrate that the proposed method can be generalized to other LMMs with alternative vision encoders, we extend our experiments to the LLaVA-1.5-7B \cite{llava1.5} model with a CLIP vision encoder \cite{CLIP}.
This LMM resizes and pads each input image to a resolution of $p \times p = 336 \times 336$ as the only input patch ($n_{\rm p} = 1$).
The number of visual features is $n_{\rm v} = 576$, and the dimension of each visual feature is $d_{\rm v} = 1024$.
The number of merged features $n_{\rm c}$ varies from $16$ to $64$, and other parameters are consistent with the setup detailed in Section \ref{sec:exp_setup}.

\begin{figure}[!t]
    \centering
    \includegraphics[width=1\linewidth]{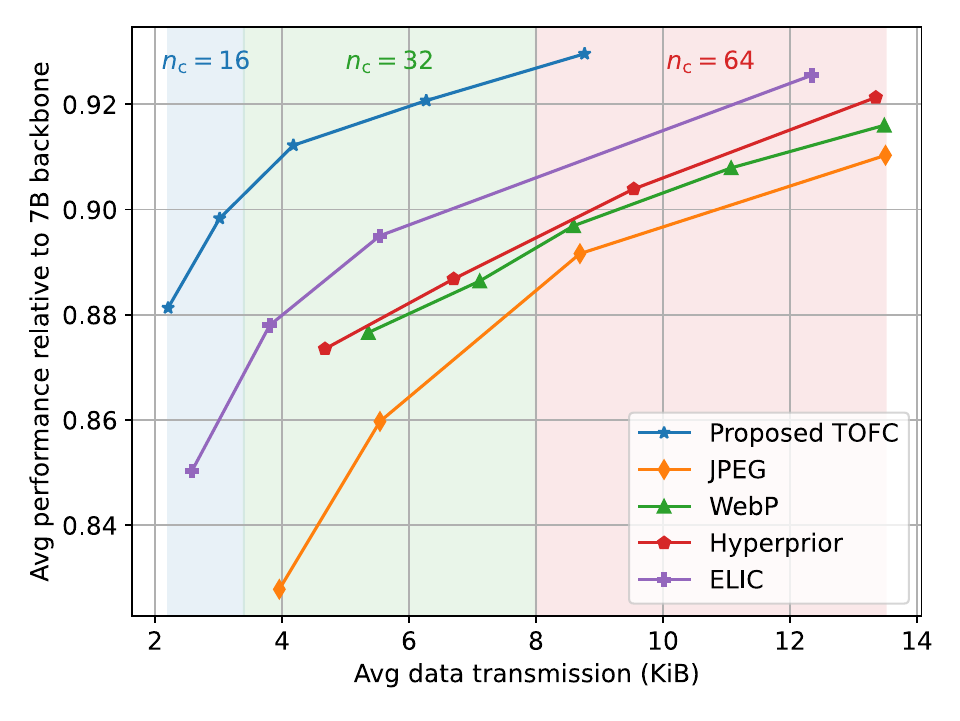}
    \caption{The rate-performance curves of different methods on the seven VQA benchmarks relative to the LLaVA-1.5-7B backbone model.}
    \label{fig_RD_avg_llava15}
\end{figure}
Fig. \ref{fig_RD_avg_llava15} illustrates the rate-performance curves of different methods on the seven VQA benchmarks relative to the LLaVA-1.5-7B backbone model.
The proposed TOFC achieves 70\%, 66\%, 63\%, and 48\% reductions in data transmission across different target performance, compared with JPEG, WebP, hyperprior, and ELIC compressions, respectively.
The data savings on LLaVA-1.5-7B model are more significant than those on LLaVA-OneVision-7B model, mainly because the input of the former model is only a single image patch.
This design reduces the number of elements in the visual features at identical number of merged features per patch $n_{\rm c}$, thus lowering the data transmission overhead of TOFC.

\begin{figure}[!t]
    \centering
    \includegraphics[width=1\linewidth]{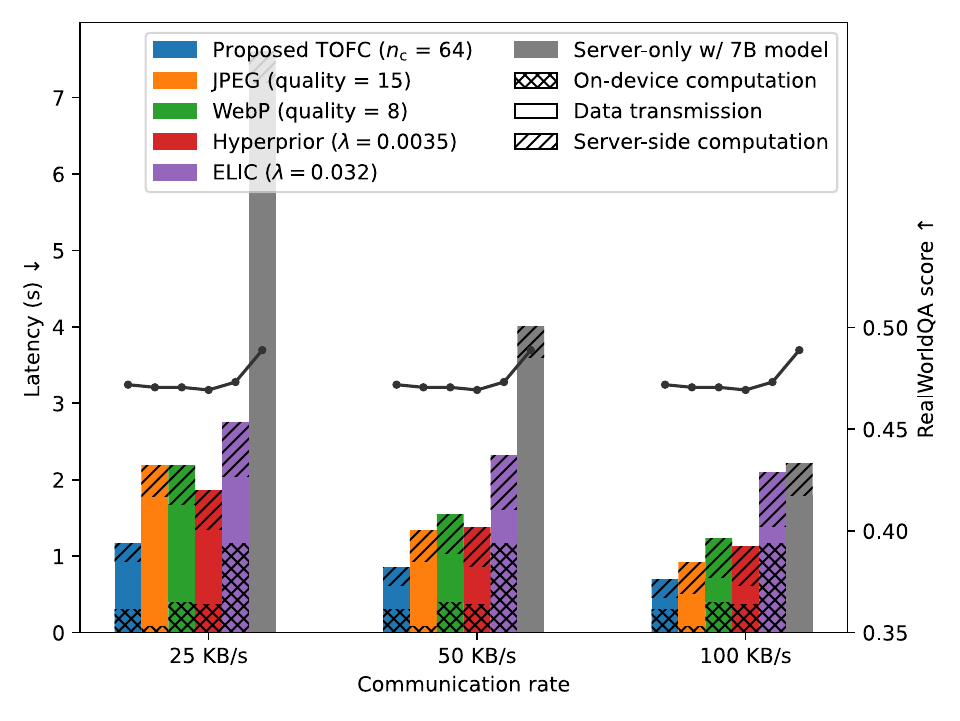}
    \caption{Average inference latency per user request (the bar plot corresponding to the left y-axis) and benchmark score (the line plot corresponding to the right y-axis) under different communication rates on RealWorldQA benchmark with the LLaVA-1.5-7B backbone model.}
    \label{fig_latency_llava15}
\end{figure}
Fig. \ref{fig_latency_llava15} shows the average inference latency per user request and the benchmark score under different communication rates on RealWorldQA benchmark with the LLaVA-1.5-7B backbone model.
The device-only inference is not incorporated as the LLaVA-1.5 series does not include a model smaller than 7B that is suitable for on-device deployment.
The proposed TOFC achieves 47\%, 47\%, 37\%, and 58\% reductions in system latency at low communication rate and similar benchmark scores, compared with JPEG, WebP, hyperprior, and ELIC compressions, respectively.
Compared to server-only inference, TOFC reduces the end-to-end latency by 68\% to 85\% across various communication rates while maintaining a 96.5\% benchmark score.
The reduction in server-side computation time is less prominent for LLaVA-1.5-7B model, whose visual tokens constitute a smaller percentage of the total input tokens, compared with LLaVA-OneVision-7B model.
However, the decreased data size of visual features minimizes the data transmission time, and the system latency reduction is thus still significant.

\subsection{Ablation Study}
\subsubsection{Feature Merging Methods}
\begin{table*}[!t]
    \centering
    \caption{Ablation Study on Feature Merging Methods\label{tab_FMM_ablation}}
    \begin{tabular}{|c|c|c|c|c|c|c|c|}
        \hline
        \textbf{Methods} & \textbf{\# tokens} & \textbf{AI2D} & \textbf{MMBench} & \textbf{MME} & \textbf{MMStar} & \textbf{RealWorldQA} & \textbf{Avg.} \\
        \hline
        Backbone model \cite{llava-ov} & 729 & 82.58 & 82.13 & 1584.6 & 61.80 & 68.89 & 100.00\% \\
        \hline
        \textbf{Proposed FMM} & 16 & \textbf{74.61} & \textbf{78.78} & \textbf{1540.8} & \textbf{50.33} & \textbf{63.92} & \textbf{91.55\%} \\
        \hline
        FMM w/o merging & 16 & 73.83 & 77.06 & 1484.4 & 48.27 & 61.18 & 88.77\% \\
        \hline
        VisionZip \cite{VisionZip} & 16 & 70.85 & 74.23 & 1484.8 & 46.86 & 51.37 & 84.05\% \\
        \hline
        VisionZip merging \cite{VisionZip} & 16 & 71.99 & 75.34 & 1525.9 & 49.55 & 57.25 & 87.70\% \\
        \hline
        Q-Former \cite{BLIP-2} & 16 & 68.69 & 68.04 & 1379.7 & 42.83 & 53.99 & 80.15\% \\
        \hline
    \end{tabular}
\end{table*}
To verify the effectiveness of the proposed FMM in compressing visual features while maintaining task performance, we conduct an ablation study on alternative feature merging methods.
Specifically, we train the backbone model separately with five different feature merging methods, using the distortion loss in (\ref{eq_D}).
The number of merged features is fixed at $n_{\rm c} = 16$ to ensure similar inference latencies across different methods.
For FMM without merging, we retain only the cluster centers selected by the DPC-KNN algorithm and discard the remaining features instead of merging them to the nearest center.
The VisionZip method \cite{VisionZip} selects four visual features that receive the highest attention and merges the unselected features into twelve features according to the algorithm in \cite{VisionZip}.
For the VisionZip merging method, we directly employ the merging algorithm proposed in \cite{VisionZip} to obtain all sixteen merged features.
The Q-Former method \cite{BLIP-2} utilizes sixteen learnable queries to extract information from key and value embeddings generated from visual features.

As shown in Table \ref{tab_FMM_ablation}, the proposed FMM outperforms alternative methods on all five benchmarks, maintaining 91.55\% of the average performance with only 2.2\% of the visual features compared with the backbone model.
The method using FMM without merging ranks second, surpassing the merging method used by VisionZip \cite{VisionZip}, which demonstrates the capability of the DPC-KNN algorithm in identifying visual features crucial for downstream tasks.
The VisionZip merging method outperforms the standard VisionZip method \cite{VisionZip}, highlighting the importance of feature merging when the number of retained features is low.
The Q-Former method \cite{BLIP-2} performs the worst, validating the superiority of training-free feature merging methods in our efficient training paradigm.

\subsubsection{Entropy Coding Methods}
\begin{figure}[!t]
    \centering
    \includegraphics[width=1\linewidth]{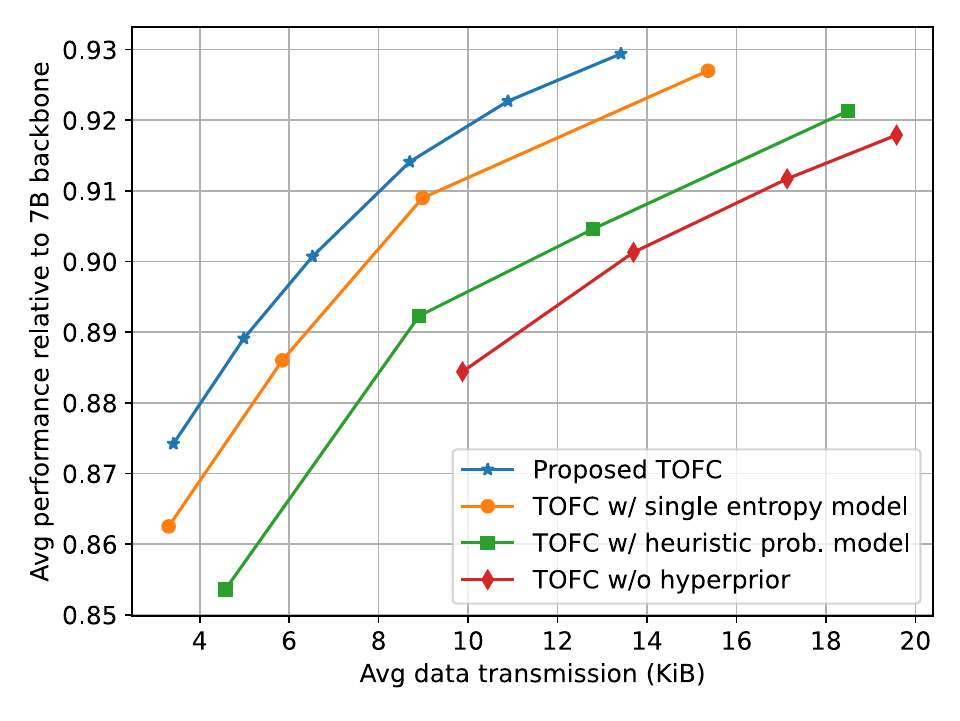}
    \caption{The rate-performance curves of the proposed TOFC method with alternative entropy coding methods.}
    \label{fig_ablation_entmodel}
\end{figure}
To verify the effectiveness of the proposed learnable and selective entropy model with hyperprior in reducing data transmission overhead while maintaining task performance, we conduct an ablation study on alternative entropy coding methods.
Specifically, we replace the entropy coding method in the proposed TOFC method with the following three methods.
First, we discard the selective entropy model architecture proposed in Section IV-C, retaining only a single entropy model.
Second, we substitute the learnable entropy model with a heuristic Laplacian probability model, whose mean and scale are calculated from the statistics of the quantized features in each batch.
Linear scaling is applied before quantization, such that the 99.9\% percentile of the merged features is mapped to the maximum integer.
Third, the learnable factorized density model, originally used for estimating the probability distribution of the hyperprior $\boldsymbol{z}$, is directly employed to model the merged features $\boldsymbol{y}$.
For the first and third learnable methods, we train multiple models with varying Lagrange multipliers $\lambda$ and numbers of merged features $n_{\rm c}$ to obtain the rate-performance curve.
For the second heuristic method, we adjust the bit length used for quantization and the number of merged features $n_{\rm c}$ to traverse the rate-performance curve.

As shown in Fig. \ref{fig_ablation_entmodel}, the proposed selective entropy model outperforms alternative entropy coding methods across all data sizes.
Compared with a single entropy model, the proposed selective entropy model achieves a 13\% to 22\% reduction in data transmission while maintaining similar performance.
Compared with the heuristic probability model, the proposed entropy model reduces data transmission overhead by approximately 40\%, showing that the learnable entropy model provides a more accurate estimation of the statistics of each merged feature.
The learnable entropy model without hyperprior delivers poor performance, confirming that the factorized density model alone is insufficient for estimating the probability distribution without the knowledge extracted by the hyperprior $\boldsymbol{z}$, which only constitutes less than 2\% of the data transmission and incurs marginal transmission overhead.

\section{Conclusions}
In this paper, we investigated a task-oriented feature compression method for multimodal understanding in a device-edge co-inference framework, where visual features are merged by clustering and encoded by a learnable and selective entropy model before feature projection.
The proposed FMM utilized DPC-KNN to reduce the number of visual features, thereby minimizing both data transmission and computational complexity.
Subsequently, we employed a learnable entropy model with hyperprior to encode and decode merged features, further reducing transmission overhead.
To improve the accuracy of the probability distribution estimated by the entropy model, multiple entropy models were adaptively selected based on the characteristics of the visual features, thus enhancing compression efficiency.
The effectiveness of the proposed TOFC method was validated through comprehensive experiments on seven VQA benchmarks.
Results demonstrated that our TOFC method achieved substantial reductions in both data transmission overhead and system latency, thus enabling delay-sensitive tasks and improving user experience.

\bibliographystyle{IEEEtran}
\bibliography{llava, NTC, TOC, token_reduction, VQAmetrics}

\begin{IEEEbiographynophoto}{Cheng Yuan}
received the B.S. degree in communication engineering from Harbin Institute of Technology, Shenzhen, China, in 2023, where he is currently pursuing the M.S. degree in electronic and information engineering. His research interests include edge AI, distributed inference, neural codecs and deep joint source-channel coding.
\end{IEEEbiographynophoto}

\begin{IEEEbiographynophoto}{Zhening Liu}
received the B.S. degree in communication engineering from Harbin Institute of Technology, Shenzhen, China, in 2023. He is currently pursuing the Ph.D. degree with the Department of Electronic and Computer Engineering, Hong Kong University of Science and Technology. His research focuses on neural data compression and generative models.
\end{IEEEbiographynophoto}

\begin{IEEEbiographynophoto}{Jiashu Lv}
received the B.S. degree in software engineering from Northwestern Polytechnical University, Xi'an, China, in 2025. He is currently pursuing the M.S. degree with the School of Software and Microelectronics, Peking University, China. His research interests include multimodal learning and model compression.
\end{IEEEbiographynophoto}

\begin{IEEEbiographynophoto}{Jiawei Shao}
is a research scientist at the Institute of Artificial Intelligence, China Telecom (TeleAI), under the direction of Prof. Xuelong Li. He is a principal investigator focusing on a wide range of topics including AI flow, large multimodal model, edge AI, task-oriented communications, and federated learning. He received his Ph.D. from the Hong Kong University of Science and Technology, and received a bachelor's degree from Beijing University of Posts and Telecommunications. He has published more than 30 research papers in top-tier journals and conferences, including Nature Communications and Nature Machine Intelligence.
\end{IEEEbiographynophoto}

\begin{IEEEbiographynophoto}{Yufei Jiang}
received the Ph.D. degree in electrical engineering and electronics from the University of Liverpool, Liverpool, U.K., in 2014. From 2014 to 2015, he was a Postdoctoral Researcher with the Department of Electrical Engineering and Electronics, University of Liverpool. From 2015 to 2017, he was a Research Associate with the Institute for Digital Communications, University of Edinburgh, Edinburgh, U.K. He is currently an Associate Professor with Harbin Institute of Technology (Shenzhen), Shenzhen, China. He has more than 100 peerreviewed publications on wireless communications and signal processing. His research interests include visible light communications, synchronization, full
duplex, age of information, URLLC, and blind source separation. Dr. Jiang received the Best Paper Award of IEEE Globecom 2019.
\end{IEEEbiographynophoto}

\begin{IEEEbiographynophoto}{Jun Zhang}
(Fellow, IEEE) received the B.Eng. degree in electronic engineering from the University of Science and Technology of China in 2004, the M.Phil. degree in information engineering from The Chinese University of Hong Kong in 2006, and the Ph.D. degree in electrical and computer engineering from The University of Texas at Austin in 2009. He is currently an Associate Professor with the Department of Electronic and Computer Engineering, The Hong Kong University of Science and Technology. He has co-authored the book Fundamentals of LTE (Prentice-Hall, 2010). His research interests include wireless communications and networking, mobile edge computing and edge AI, and integrated AI and communications. He is an IEEE ComSoc Distinguished Lecturer. He was a co-recipient of several best paper awards, including the 2021 Best Survey Paper Award of the IEEE Communications Society, the 2019 IEEE Communications Society \& Information Theory Society Joint Paper Award, and the 2016 Marconi Prize Paper Award in Wireless Communications. Two papers he has co-authored received the Young Author Best Paper Award of the IEEE Signal Processing Society in 2016 and 2018, respectively. He also received the 2016 IEEE ComSoc Asia-Pacific Best Young Researcher Award. He is an Area Editor of IEEE Transactions on Wireless Communications and IEEE Transactions on Machine Learning in Communications and Networking.
\end{IEEEbiographynophoto}

\begin{IEEEbiographynophoto}{Xuelong Li}
(M‘02-SM’07-F‘12) is the CTO and Chief Scientist of China Telecom, where he founded the Institute of Artificial Intelligence (TeleAI) of China Telecom.
\end{IEEEbiographynophoto}






\vfill

\end{document}